%% file: Qualitative_Describing_Function.tex
\documentclass[letterpaper, 10 pt, conference]{ieeeconf}

\IEEEoverridecommandlockouts

\usepackage{cite}
\usepackage{amsmath,amssymb,amsfonts}
\usepackage{algorithmic}
\usepackage{graphicx}
\usepackage{textcomp}
\usepackage{xcolor}
\usepackage[switch]{lineno}
\usepackage{algorithmic}
\usepackage{algorithm}

\usepackage{orcidlink}
 \usepackage[italian]{babel}
 \usepackage{pstricks}
 \usepackage{verbatim}
 \usepackage{graphicx}
 \usepackage{psfrag}
 \usepackage{fancybox}
 \usepackage{amsmath}
 \usepackage{amsfonts}
 \usepackage{amssymb}
 \usepackage{listings}
 \usepackage{enumerate}
 \usepackage{xurl}

\usepackage{hyperref}

 \makeatletter
 \newcommand\notsotiny{\@setfontsize\notsotiny\@vipt\@viipt}
 \makeatother

\input{pst-plot}
\input{pst-node}
\input{pst-coil}

\input{mathbase.tex}

\definecolor{mygreen}{RGB}{28,172,0}

 \newtheorem{Assumption}{\bf Assumption}

 \newtheorem{Prop}{\bf Property}

 \newtheorem{Remar}{\bf Remark}

\newcommand{\ENG}{1=1}

\newcommand{\ItaEng}[2]{\ifnum\ENG\relax#2\else#1\fi}

\definecolor{mio_red}{rgb}{0.0, 0.0, 0.0}
\definecolor{mioo_red}{rgb}{0.0, 0.0, 0.0}

\title{\LARGE \bf
An Approach for the Qualitative Graphical Representation of the Describing Function in Nonlinear Systems Stability Analysis
}

\author{Davide Tebaldi\orcidlink{0000-0003-1432-0489} \IEEEmembership{Member, IEEE}  and Roberto Zanasi\orcidlink{0000-0001-5507-825X}
\thanks{The work was partly supported by the University of Modena and Reggio Emilia
through the action FARD (Finanziamento Ateneo Ricerca Dipartimentale) 2023/2024, and funded under the National Recovery and Resilience Plan (NRRP), Mission 04 Component 2 Investment 1.5 - NextGenerationEU, Call for tender n. 3277 dated 30/12/2021
Award Number:  0001052 dated 23/06/2022.}
\thanks{The authors are
with the Department of Engineering ``Enzo Ferrari'', University of
Modena and Reggio Emilia, Modena, Via Pietro Vivarelli 10,
41125 Modena, Italy. E-mails: \{davide.tebaldi, roberto.zanasi\}@unimore.it.
}
\thanks{This work has been submitted to the IEEE for possible publication.
Copyright may be transferred without notice, after which this version may
no longer be accessible.}
}

\begin{document}

\maketitle
\thispagestyle{empty}
\pagestyle{empty}

\begin{abstract}

The describing function method is a useful tool for the qualitative analysis of limit cycles in the stability analysis of nonlinear systems. This method is inherently approximate; therefore, it should be used for a fast qualitative analysis of the considered systems. However, plotting the exact describing function requires heavy mathematical calculations, reducing interest in this method especially from the point of view of control education. The objective of this paper is to enhance the describing function method by
providing a new approach for the qualitative plotting of the describing function for piecewise nonlinearities involving discontinuities.
Unlike the standard method,
the proposed approach allows for a straightforward, hand-drawn plotting of the describing function using the rules introduced in this paper, simply by analyzing the shape of the nonlinearity.
The proposed case studies show that the limit cycles estimation performed using the standard exact plotting of the describing function yields the same qualitative results as those obtained using the proposed qualitative method for plotting the describing function.

\end{abstract}

\section{INTRODUCTION}
Persistent oscillatory behaviors, typically referred to as limit cycles, are an inherent property of a large variety of systems of different nature.
Limit cycles are indeed present in engineering systems belonging to different energetic domains, including
mechanical systems \cite{10.1115/1.4035190} such as robotic systems, electrical systems \cite{kyurkchiev2022extended} such as electrical oscillators, and hydraulic systems \cite{9385620} such as hydraulic actuators.

\begin{figure}[t]
\centering
\thicklines
\setlength{\unitlength}{2.68mm}
\psset{unit=1.0\unitlength}
\begin{pspicture}(10,1)(33,8)
\put(13,3){\framebox(6,4){
\begin{pspicture}(-3,-2)(3,2)
\pscurve(-2.5,-1.2)(-1,-1)(1,1)(2.5,1.2)
\psline[linewidth=0.4pt]{->}(-2.5,0)(2.5,0)
\psline[linewidth=0.4pt]{->}(0,-1.5)(0,1.5)
\rput[r](-0.125,1.125){\small $y(x)$}
\rput[t](2.125,-0.25){\small $x$}
\end{pspicture}
}}
\psline{->}(19,5)(25,5)
\put(25,3){\framebox(6,4){$-G(s)$}}
\psline{->}(31,5)(34.8,5)
\psline{->}(34.8,5)(34.8,0.5)(9.2,0.5)(9.2,5)(13,5)
\rput[lb](10.5,5.25){\small $x(t)$}
\rput[lb](21.25,5.25){\small $y(t)$}
\rput[lb](14.68,1.35){\small $F(X)$}
\end{pspicture}
\vspace{-1.1mm}
\caption{Considered nonlinear autonomous feedback system.}
\label{nonlinear_feedback_structure}
\vspace{-3mm}
\end{figure}
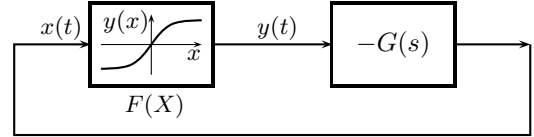

One of the most effective tools to study the existence and the characteristics of limit cycles is the describing function method
\cite{desfcn2}.
With reference to the system of Fig.~\ref{nonlinear_feedback_structure}, limit cycles can exist under certain assumptions if the so-called self-sustaining equation
$F(X)\,G(j\omega)=-1$
exhibits solutions, where $G(j\omega)$ is the frequency response function of the linear  system $G(s)$, and $F(X)$ is the describing function of the considered nonlinearity.
Despite being non-rigorous, the describing  function method finds several applications,
including photovoltaic systems \cite{zhang2018power}, disturbance rejection in nonlinear active disturbance rejection control
\cite{6213547}, and the identification of the structural nonlinearities in multi-degree-of-freedom nonlinear systems \cite{altro1}.

Through the years, many different modifications or additions to the describing function method have been made. In \cite{1085408}, a stability criterion is provided, giving an interesting justification of the quasi-static stability criterion
upon which the describing function method is based. The describing function method is studied in \cite{244905} for limit cycles analysis
in switched circuits, while a refinement of the approximated solution given by this method is proposed in \cite{1656433} and an extension
to systems exhibiting more than one nonlinearity in given in \cite{inproceedings81}.
Thanks to computer aided control system design software, interactive tools have been developed for limit cycles analysis
using the describing function method~\cite{DORMIDO2024364}.

However, the use of graphical tools such as
the fast hand-drawing of the describing function can be an important ability for control engineers. Indeed, despite being an effective method, the main limitation associated with the describing function, especially from the point of view of control education, is the derivation of the analytical expression of the describing function, which often requires a detailed analysis of the considered nonlinearity and is derived using heavy mathematical calculations \cite{1104997}.
Considering that the describing function method is non-rigorous, since it is based on an approximated analysis \cite{1047017}, the complexity of calculation diminishes the easiness and interest of application of this method.

The objective of this paper is therefore to enhance the standard describing function method by providing an approach for the approximate hand-drawing of the describing function itself, with no need to perform complex calculations, making the method more appealing from an educational point of view.
The proposed method applies to piecewise nonlinearities involving discontinuities, and consists in a series of rules for drawing the qualitative describing function by hand very quickly, by simply looking at the nonlinear characteristic. Furthermore, the proposed method is also translated into an algorithm involving simpler calculations than the standard complex analytical expression of the describing function $F(X)$.
The considered case studies show that the analysis and limit cycles estimation carried out using the qualitative describing function plotting give, qualitatively, the same results as those obtained using the standard describing function plotting.

The remainder of this paper is structured as follows. The describing function of two standard nonlinearities are recalled in Sec.~\ref{Section_one_sect1}, while an algorithm for plotting the describing function of piecewise nonlinearities with discontinuities is given in Sec.~\ref{Section_piecewise}. The new approach for the qualitative plotting of the describing function is proposed in Sec.~\ref{Section_one_sect3}, and applied to the limit cycle estimation of two case studies in Sec.~\ref{Section_one_sect4}. The conclusions are given in Sec.~\ref{conclusions_sect}.


\section{Describing function}\label{Section_one_sect1}

Let $x(t)\!=\!X\,\sin(\omega t)$  be the sinusoidal input of the nonlinear function $y(x)$ shown in Fig.~\ref{nonlinear_feedback_structure}.
The corresponding output signal  $y(t)$ is a periodic signal
that can be expanded in a Fourier series as follows:
\begin{equation}
y(t)=\sum_{n=1}^{\infty} Y_n\,\sin\,(n\omega t+\varphi_n).
 \label{yt_Fourier}
\end{equation}
The constant term $Y_0$ is missing in \eqref{yt_Fourier} due to Assumption~\ref{Assumption1}.

\begin{Assumption} \label{Assumption1} Reference is made to the nonlinear feedback structure shown in Fig.~\ref{nonlinear_feedback_structure}, where
the nonlinear function $y(x)$ is purely algebraic, symmetric with respect to the origin, i.e. $y(-x)\!=\!-y(x)$, and independent of the frequency $\omega$ of the input signal $x(t)$.
\end{Assumption}

The use of the describing function is based on the following approximation: all the higher order harmonics (i.e. for $n\geq2$) in the Fourier series \eqref{yt_Fourier} are neglected, while the first harmonic $y(t)\simeq Y_1(X)\ \sin\,\left(\omega t+\varphi_1(X)\right)$ is considered.

The describing function $F(X)$ of the  nonlinear function $y(x)$
links the  sinusoidal input $x(t)\!=\!X\,\sin(\omega t)$  to the first harmonic $Y_1(X)\ \sin\,\left(\omega t+\varphi_1(X)\right)$ of the output signal $y(t)$~\cite{desfcn2} and, in the general case, is defined as follows:
\begin{equation}
  F(X)
  =  \frac{Y_1(X)}{X}\,\,e^{j\varphi_1(X)},
 \label{FdiX}
\end{equation}
where $ \ts Y_1(X)=\sqrt{a_1^{2}(X)+b_1^{2}(X)}$, $\ts \varphi_1(X)=\arctan\, \frac{a_1(X)}{b_1(X)}$,
\begin{equation}
\begin{array}{l}
   \ts a_1(X)=\frac{1}{\pi}\!\int_{-\pi}^{\pi} \!y(X\sin\theta)\,\cos\theta\, d\theta,
     \\[2mm]
         \ts  b_1(X)=\frac{1}{\pi}\!\int_{-\pi}^{\pi}
\!y(X\sin\theta)\,\sin\theta\, d\theta,
\end{array}
 \label{a1b1}
\end{equation}
and $\theta=\omega t$.
Due to the symmetry hypothesis in Assumption~\ref{Assumption1}, the coefficient $a_1(X)$ in \eqref{a1b1} is zero,
and the describing function $F(X)$
simplifies as follows~\cite{desfcn2}:
\begin{equation}
  F(X)
  =  \frac{b_1(X)}{X}
  =\frac{4}{\pi\,X}\!\int_{0}^{\frac{\pi}{2}}
\!y(X\sin\theta)\,\sin\theta\, d\theta.
 \label{FdiX_b1}
\end{equation}

\subsubsection{Dead Zone}\label{Section_y1x}
 The describing function $F_d\!\left(\!\ts m,X_1,X\!\right)$ of the dead zone $y_d(x)=y_d\!\left(\!\ts m_1,\!\ts X_1,x\!\right) $ shown in Fig.~\ref{fig:y2xd}(a) is the following~\cite{1104997}:
\begin{equation}
 F_d(X) \!=\! F_d\!\left(\!\ts m,X_1,X\!\right)
 \!=\!\left\{
\begin{array}{@{}lcl@{}}
 0 &\mbox{if} & X\!<\!X_1,
  \\
  m \,\Phi\!\left(\! \frac{X}{X_1}\right)\!
  &\mbox{if} & X\!\ge\!X_1,
   \end{array}
 \right.
 \label{FX_of_dead_zone}
\end{equation}
where:
\begin{equation}
\ts
\Phi\!\left(\!\frac{X}{X_1}\!\right)
=
1-\frac{2}{\pi}
\left(\arcsin\frac{X_1}{X}\!+\!\frac{X_1}{X}
\sqrt{1\!-\!\left(\frac{X_1}{X}\right)^2}\right).
 \label{Phi_function}
\end{equation}
The shape of function $F_d(X)$ in \eqref{FX_of_dead_zone} is shown in Fig.~\ref{fig:y2xd}(b).  The derivative of function $F_{d}(X)$ with respect to $X$ is the following:
\[
\frac{F_{d}(X)}{dX}=\frac{4 m X_1 (X^2-X_1^2)}{\pi X^3\sqrt{X^2-X_1^2}},
\]
from which it follows that
the slope of function $F_{d}(X)$ is zero for $X=X_1$ and is positive for $X>X_1$. This means that function $F_{d}(X)$ always increases for $X>X_1$ and reaches its maximum value $m$ when $X\rightarrow\infty$.

\begin{figure}
\centering
\setlength{\unitlength}{1.8mm}
\psset{unit=\unitlength}
\SpecialCoor
\begin{pspicture}(-6,-2)(14,9.5)
\footnotesize
\psline[linewidth=0.4pt]{->}(-2,0)(14,0)
\psline[linewidth=0.4pt]{->}(0,-2)(0,9.5)
\psline[linewidth=0.8pt]{-}(0,0)(5,0)
\rput[t](5,-0.5){$X_1$}
\psline[linewidth=0.2pt,linestyle=dashed]{-}(5,0)(5,0)
\psline[linewidth=0.8pt]{-}(5,0)(12,7)
\rput[t](13,-0.5){$x$}
\rput[r](-0.25,8){$y_d(x)$}
\psline[linewidth=0.4pt]{-}(7,2)(9,2)(9,4)
\rput[l](9.5,3){$m$}
\rput(5.6,8){(a)}
\end{pspicture}
\hspace{1.0cm}
\begin{pspicture}(-6,-2)(14,9.5)
\footnotesize
\psline[linewidth=0.4pt]{->}(-2,0)(14,0)
\psline[linewidth=0.4pt]{->}(0,-2)(0,9.5)
\psline[linewidth=0.8pt]{-}(0,0)(5,0)
\rput[t](5,-0.5){\scriptsize $X_1$}
\psline[linewidth=0.2pt,linestyle=dashed]{-}(5,0)(5,0)
\psbezier[linewidth=0.8pt](5,0)(5.5,2)(6,4)(13,4)
\psline[linewidth=0.2pt,linestyle=dashed]{-}(0,4)(13,4)
\rput[r](-0.5,4){$m$}
\rput[t](13,-0.5){$X$}
\rput[r](-0.25,8){$F_d(X)$}
\rput(5.6,8.5){(b)}
\end{pspicture}
\caption{(a) Dead zone function $y_d(x)$; (b) Describing function $F_d(X)$.} \label{fig:y2xd}
\end{figure}
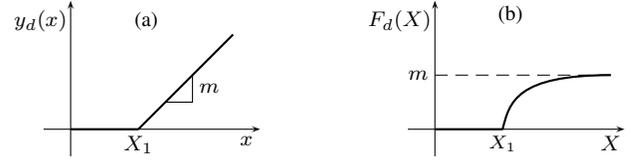

\begin{figure}
\centering
\setlength{\unitlength}{1.8mm}
\psset{unit=\unitlength}
\SpecialCoor
\begin{pspicture}(-6,-2)(14,9.5)
\footnotesize
\psline[linewidth=0.4pt]{->}(-2,0)(14,0)
\psline[linewidth=0.4pt]{->}(0,-2)(0,9.5)
\psline[linewidth=0.8pt]{-}(0,0)(5,0)
\psline[linewidth=0.2pt,linestyle=dashed]{-}(5,0)(5,0)
\psline[linewidth=0.8pt]{-}(5,0)(5,5)
\rput[t](5,-0.5){$X_1$}
\psline[linewidth=0.2pt,linestyle=dashed]{-}(5,0)(5,5)
\rput[r](-0.25,5){$Y_1$}
\psline[linewidth=0.2pt,linestyle=dashed]{-}(0,5)(5,5)
\psline[linewidth=0.8pt]{-}(5,5)(13,5)
\rput[t](13,-0.5){$x$}
\rput[r](-0.25,8){$y_r(x)$}
\rput(5.6,8){(a)}
\end{pspicture}
\hspace{1.0cm}
\begin{pspicture}(-6,-2)(14,9.5)
\footnotesize
\psline[linewidth=0.4pt]{->}(-2,0)(14,0)
\psline[linewidth=0.4pt]{->}(0,-2)(0,9.5)
\psline[linewidth=0.8pt]{-}(0,0)(5,0)
\psline[linewidth=0.2pt,linestyle=dashed]{-}(5,0)(5,0)
\rput[tr](5.7,-0.5){\scriptsize $X_1$}
\psbezier[linewidth=0.8pt](5,0)(6.5,10)(7,0)(13,0)
\psline[linewidth=0.2pt,linestyle=dashed]{-}(6.5,0)(6.5,4.5)
\psline[linewidth=0.2pt,linestyle=dashed]{-}(0,4.5)(6.5,4.5)
\rput[tl](6.2,-0.4){\scriptsize$X_r^*$}
\rput[r](-0.5,4.5){$F_r^*$}
\rput[t](13,-0.5){$X$}
\rput[r](-0.25,8){$F_r(X)$}
\rput(5.6,8.5){(b)}
\end{pspicture}
\caption{(a) Relay  with threshold $y_r(x)$; (b) Describing function $F_r(X)$.}
   \vspace{-3mm}
\label{fig:y2xr}
\end{figure}
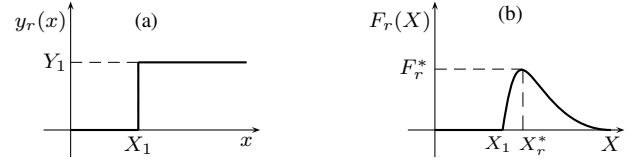

\subsubsection{Relay with threshold}\label{Section_y1x}
The describing function $F_r\!\left(\!\ts Y_1,X_1,X\!\right)$ of the relay with threshold $y_r(x)\!=\!y_r(Y_1,X_1,x)$ shown in Fig.~\ref{fig:y2xr}(a) is~\cite{1104997}:

\vspace{-5mm}
\begin{equation}
 F_r(X)\!=\!
F_r\!\left(\!\ts Y_1,X_1,X\!\right) \!=\!\left\{
\begin{array}{@{}l@{\;}c@{\;}l@{}}
 0 &\mbox{if} & X\!<\!X_1,
  \\
Y_1\Psi\!\left(\!X_1,X\!\right)
  &\mbox{if} & X\!\ge\!X_1,
   \end{array}
 \right.
 \label{FX_of_rele_with_dead_zone}
\end{equation}
where:
\begin{equation}
\ts
\Psi\!\left(\!X_1,X\!\right)
=
\frac{4}{\pi X}\sqrt{1\!-\!\left(\frac{X_1}{X}
  \right)^2}.
 \label{Psi_function}
\end{equation}
 The shape of function $F_r(X)$ is shown in Fig.~\ref{fig:y2xr}(b).  The derivative of function $F_{r}(X)$ with respect to $X$ is:
\[\frac{F_{r}(X)}{dX}=\frac{4 Y_1 (2 X_1^2-X^2)}{\pi X^3\sqrt{X^2-X_1^2}},
\]
from which it follows that
the slope of function $F_{r}(X)$ is infinite for $X =X_1$. The maximum $F_r^*$ of function $F_{r}(X)$ occurs at $X=X_r^*=\sqrt{2}X_1$ and is given by $F_r^*=\frac{2Y_1}{\pi X_1}$.

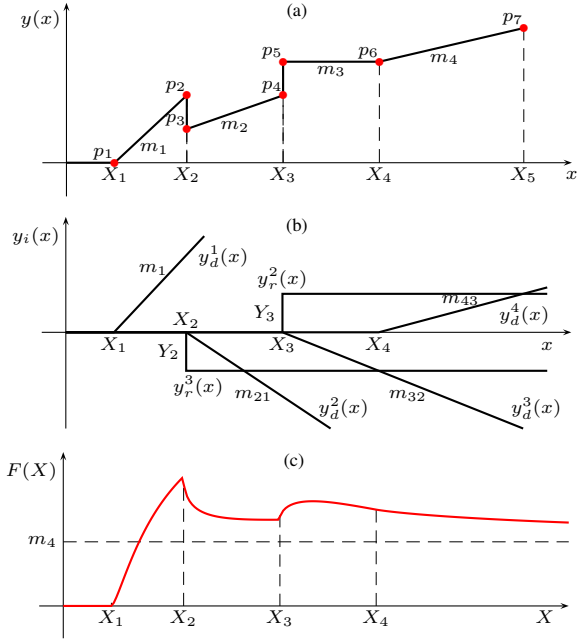
\begin{figure}[tb]
    \centering
{ \scriptsize
\setlength{\unitlength}{3.2mm}
\psset{unit=\unitlength}
\psset{xunit=\unitlength}
\psset{yunit=0.7\unitlength}
\SpecialCoor
\begin{pspicture}(-2,-1.5)(21,9.5)
\psline[linewidth=0.4pt]{->}(-1,0)(21,0)
\psline[linewidth=0.4pt]{->}(0,-2)(0,9.5)
\psline[linewidth=0.8pt]{-}(0,0)(2,0)
\rput[t](2,-0.25){$X_{1}$}
\psline[linewidth=0.2pt,linestyle=dashed]{-}(2,0)(2,0)
\psline[linewidth=0.8pt]{-}(2,0)(5,4)
\rput[t](5,-0.25){$X_{2}$}
\rput[lt](3.05,1.1625){$m_{1}$}
\psline[linewidth=0.2pt,linestyle=dashed]{-}(5,0)(5,4)
\psline[linewidth=0.8pt]{-}(5,4)(5,2)
\psline[linewidth=0.2pt,linestyle=dashed]{-}(5,0)(5,2)
\psline[linewidth=0.8pt]{-}(5,2)(9,4)
\rput[t](9,-0.25){$X_{3}$}
\rput[lt](6.4,2.4625){$m_{2}$}
\psline[linewidth=0.2pt,linestyle=dashed]{-}(9,0)(9,4)
\psline[linewidth=0.8pt]{-}(9,4)(9,6)
\psline[linewidth=0.2pt,linestyle=dashed]{-}(9,0)(9,6)
\psline[linewidth=0.8pt]{-}(9,6)(13,6)
\rput[t](13,-0.25){$X_{4}$}
\rput[lt](10.4,5.7625){$m_{3}$}
\psline[linewidth=0.2pt,linestyle=dashed]{-}(13,0)(13,6)
\psline[linewidth=0.8pt]{-}(13,6)(19,8)
\rput[t](19,-0.25){$X_{5}$}
\rput[lt](15.1,6.4625){$m_{4}$}
\psline[linewidth=0.2pt,linestyle=dashed]{-}(19,0)(19,8)
\psdot[linecolor=red, dotsize=3pt](2,0)
\rput[br](2,0.2){$p_1$}
\psdot[linecolor=red, dotsize=3pt](5,4)
\rput[br](5,4.2){$p_2$}
\psdot[linecolor=red, dotsize=3pt](5,2)
\rput[br](5,2.2){$p_3$}
\psdot[linecolor=red, dotsize=3pt](9,4)
\rput[br](9,4.2){$p_4$}
\psdot[linecolor=red, dotsize=3pt](9,6)
\rput[br](9,6.2){$p_5$}
\psdot[linecolor=red, dotsize=3pt](13,6)
\rput[br](13,6.2){$p_6$}
\psdot[linecolor=red, dotsize=3pt](19,8)
\rput[br](19,8.2){$p_7$}
\rput[l](9.16,9.0){\scriptsize (a)}
\rput[t](21,-0.5){$x$}
\rput[r](-0.25,8.5){$y(x)$}
\end{pspicture}

\psset{yunit=0.8\unitlength}
\;\begin{pspicture}(-2,-4.55)(21,7.5)
\psline[linewidth=0.4pt]{->}(-1,0)(21,0)
\psline[linewidth=0.4pt]{->}(0,-5)(0,6)
\psline[linewidth=0.8pt,linecolor=black]{-}(5,0)(5,-2)(20,-2)
\rput[r](4.75,-1){$Y_2$}
\psline[linewidth=0.8pt,linecolor=black]{-}(9,0)(9,2)(20,2)
\rput[r](8.75,1){$Y_3$}
\psline[linewidth=0.8pt,linecolor=black]{-}(0,0)(2,0)(5.75,5)
\rput[rb](4.25,3){$m_{1}$}
\psline[linewidth=0.8pt,linecolor=black]{-}(0,0)(5,0)(11,-5)
\rput[rt](8.6,-3){$m_{21}$}
\psline[linewidth=0.8pt,linecolor=black]{-}(0,0)(9,0)(19,-5)
\rput[rt](15,-3){$m_{32}$}
\psline[linewidth=0.8pt,linecolor=black]{-}(0,0)(13,0)(20,2.3333)
\rput[rb](17.2,1.4){$m_{43}$}
\rput[l](9.16,5.5){\scriptsize (b)}
\rput[t](20,-0.5){$x$}
\rput[r](-0.25,5){$y_i(x)$}
\rput[t](2,-0.25){$X_{1}$}
\rput[b](5, 0.25){$X_{2}$}
\rput[t](9,-0.25){$X_{3}$}
\rput[t](13,-0.25){$X_{4}$}
\rput[tl](5.5,4.5){$y_d^1(x)$}
\rput[bl](10.5,-4.5){$y_d^2(x)$}
\rput[bl](18.5,-4.5){$y_d^3(x)$}
\rput[tl](18.0,1.5){$y_d^4(x)$}
\rput[b](9.0,2.25){$y_r^2(x)$}
\rput[t](5.5,-2.25){$y_r^3(x)$}
\end{pspicture}
\vspace{3mm}

\psset{xunit=\unitlength}
\psset{yunit=8\unitlength}
\begin{pspicture}(-2,-0.1)(21,0.85)
\psline[linewidth=0.4pt]{->}(-1,0)(21,0)
\psline[linewidth=0.4pt]{->}(0,-0.168)(0,0.8)
\psline[linewidth=0.1pt,linestyle=dashed](2,0)(2,0.016513)
\rput[t](2,-0.025){$X_1$}
\psline[linewidth=0.1pt,linestyle=dashed](5,0)(5,0.61355)
\rput[t](5,-0.025){$X_2$}
\psline[linewidth=0.1pt,linestyle=dashed](9,0)(9,0.47105)
\rput[t](9,-0.025){$X_3$}
\psline[linewidth=0.1pt,linestyle=dashed](13,0)(13,0.50017)
\rput[t](13,-0.025){$X_4$}
\psline[linewidth=0.1pt,linestyle=dashed](0,0.33333)(21,0.33333)
\rput[r](-0.25,0.33333){$m_4$}
\psline[linewidth=0.8pt,linestyle=solid,linecolor=red](0,0)(0.105,0)(0.21,0)(0.315,0)(0.42,0)(0.525,0)(0.63,0)(0.735,0)(0.84,0)(0.945,0)(1.05,0)(1.155,0)(1.26,0)(1.365,0)(1.47,0)(1.575,0)(1.68,0)(1.785,0)(1.89,0)(1.995,0)(2.1,0.016513)(2.205,0.044734)(2.31,0.077083)(2.415,0.11103)(2.52,0.1453)(2.625,0.17916)(2.73,0.21222)(2.835,0.24422)(2.94,0.27506)(3.045,0.30468)(3.15,0.33305)(3.255,0.3602)(3.36,0.38616)(3.465,0.41096)(3.57,0.43467)(3.675,0.45732)(3.78,0.47898)(3.885,0.49969)(3.99,0.51949)(4.095,0.53846)(4.2,0.55661)(4.305,0.57401)(4.41,0.5907)(4.515,0.6067)(4.62,0.62207)(4.725,0.63682)(4.83,0.65101)(4.935,0.66466)(5.04,0.61355)(5.145,0.56904)(5.25,0.5444)(5.355,0.52719)(5.46,0.51416)(5.565,0.50386)(5.67,0.49548)(5.775,0.48855)(5.88,0.48274)(5.985,0.47781)(6.09,0.4736)(6.195,0.46999)(6.3,0.46687)(6.405,0.46417)(6.51,0.46183)(6.615,0.45979)(6.72,0.45802)(6.825,0.45647)(6.93,0.45513)(7.035,0.45396)(7.14,0.45294)(7.245,0.45206)(7.35,0.4513)(7.455,0.45064)(7.56,0.45009)(7.665,0.44961)(7.77,0.44922)(7.875,0.44889)(7.98,0.44862)(8.085,0.4484)(8.19,0.44824)(8.295,0.44812)(8.4,0.44804)(8.505,0.44799)(8.61,0.44798)(8.715,0.44799)(8.82,0.44804)(8.925,0.4481)(9.03,0.47105)(9.135,0.49497)(9.24,0.50833)(9.345,0.51769)(9.45,0.52469)(9.555,0.53007)(9.66,0.53423)(9.765,0.53743)(9.87,0.53987)(9.975,0.54167)(10.08,0.54294)(10.185,0.54375)(10.29,0.54416)(10.395,0.54423)(10.5,0.54401)(10.605,0.54352)(10.71,0.5428)(10.815,0.54188)(10.92,0.54078)(11.025,0.53951)(11.13,0.5381)(11.235,0.53657)(11.34,0.53492)(11.445,0.53317)(11.55,0.53133)(11.655,0.52941)(11.76,0.52741)(11.865,0.52536)(11.97,0.52324)(12.075,0.52108)(12.18,0.51887)(12.285,0.51662)(12.39,0.51434)(12.495,0.51203)(12.6,0.50969)(12.705,0.50733)(12.81,0.50495)(12.915,0.50256)(13.02,0.50017)(13.125,0.4981)(13.23,0.49622)(13.335,0.49446)(13.44,0.4928)(13.545,0.49121)(13.65,0.48969)(13.755,0.48823)(13.86,0.48682)(13.965,0.48545)(14.07,0.48413)(14.175,0.48285)(14.28,0.4816)(14.385,0.48038)(14.49,0.47919)(14.595,0.47803)(14.7,0.4769)(14.805,0.47579)(14.91,0.47471)(15.015,0.47365)(15.12,0.47261)(15.225,0.47159)(15.33,0.47059)(15.435,0.4696)(15.54,0.46864)(15.645,0.46769)(15.75,0.46676)(15.855,0.46584)(15.96,0.46494)(16.065,0.46405)(16.17,0.46318)(16.275,0.46232)(16.38,0.46148)(16.485,0.46065)(16.59,0.45982)(16.695,0.45902)(16.8,0.45822)(16.905,0.45743)(17.01,0.45666)(17.115,0.45589)(17.22,0.45514)(17.325,0.4544)(17.43,0.45366)(17.535,0.45294)(17.64,0.45223)(17.745,0.45152)(17.85,0.45083)(17.955,0.45014)(18.06,0.44946)(18.165,0.44879)(18.27,0.44813)(18.375,0.44747)(18.48,0.44683)(18.585,0.44619)(18.69,0.44556)(18.795,0.44493)(18.9,0.44432)(19.005,0.44371)(19.11,0.4431)(19.215,0.44251)(19.32,0.44192)(19.425,0.44134)(19.53,0.44076)(19.635,0.44019)(19.74,0.43963)(19.845,0.43907)(19.95,0.43852)(20.055,0.43798)(20.16,0.43744)(20.265,0.4369)(20.37,0.43637)(20.475,0.43585)(20.58,0.43533)(20.685,0.43482)(20.79,0.43432)(20.895,0.43381)(21,0.43332)
\rput[l](9.16,0.75){\scriptsize (c)}
\rput[t](20,-0.02){$X$}
\rput[r](-0.25,0.7){$F(X)$}
\end{pspicture}
 } %
 \caption{(a) Piecewise nonlinear function $y(x)$; (b) Components of $y(x)$ in \eqref{yx_d_r}; (c) Describing function  $F(X)$ in \eqref{yx_d_r_bis} computed using Algorithm~\ref{Algorithm_optimal_eff}.}
   \vspace{-3mm}
    \label{fig:piecewise_nonlinear}
\end{figure}

\section{Describing function $F(X)$ of piecewise nonlinear functions $y(x)$}\label{Section_piecewise}
The describing functions of several classes of
nonlinearities have been determined \cite{1104997}. In this paper, we focus on
the describing function $F(X)$ of a generic piecewise nonlinear function $y(x)$.

\begin{Prop}
The describing function $F(X)$ of a piecewise nonlinear function $y(x)$ with discontinuities
can always be obtained as a linear combination  of a certain number of describing functions $F_d(X)$ and  $F_r(X)$ of the two nonlinear functions $y_{d}(x)$ and  $y_{r}(x)$ shown in Fig.~\ref{fig:y2xd} and Fig.~\ref{fig:y2xr}.
\label{Prop_1}
\end{Prop}
\vspace{1mm}
An an example, reference is made to the case study of Fig.~\ref{fig:piecewise_nonlinear}(a). In this case, the function $y(x)$ can be obtained  as the sum of six functions of type $y_{d}(x)$ and $y_{r}(x)$, as in Fig.~\ref{fig:piecewise_nonlinear}(b):
\begin{equation}
\begin{array}{r@{\,}c@{\,}l}
y(x) &=&
y_d^1\!\left(\!m_1,\!X_1,\!x\!\right)
\!+\!y_d^2\!\left(\!m_{21},\!X_2,\!x\!\right)
\!+\! y_r^2\!\left(\!Y_2,\!X_2,x\!\right)
\\[1mm]
 &&
\!\!\!\!+ y_d^3\!\left(\!m_{32},\!X_3,\!x\!\right)
\!+\! y_r^3\!\left(\!Y_3,\!X_3,\!x\!\right)
\!+\! y_d^4\!\left(\!m_{43},\!X_4,\!x\!\right),
\end{array}
\label{yx_d_r}
\end{equation}
where $m_{21}\!=\!m_2\!-\!m_1$,
$m_{32}\!=\!m_3\!-\!m_2\!=\!-m_2$ and
$m_{43}\!=\!m_4\!-\!m_3=m_4$.
From \eqref{yx_d_r}, it follows that the corresponding describing function $F(X)$,
shown in Fig.~\ref{fig:piecewise_nonlinear}(c),
is the sum of
six  describing functions of type $F_{d}(X)$ and $F_{r}(X)$:
\begin{equation}
\begin{array}{@{\!\!\!\!}r@{}c@{}l@{}}
F(X) &=& \ts
F_d\!\left(\!\ts m_1,\!X_1,\!X\!\right)
\!+\! F_d\!\left(\!\ts m_{21},\!X_2,\!X\!\right)
\!+\! F_r\!\left(\!\ts Y_1,\!X_2,\!X\!\right)
\\[1mm]
 &&
\!\!\!\!\!\!\!+F_d\!\left(\!\ts m_{32},\!X_3,\!X\!\right)
\!+\! F_r\!\left(\!\ts Y_3,\!X_3,\!X\!\right)
\!+\! F_d\!\left(\!\ts m_{43},\!X_4,\!X\!\right).
\end{array}\!
\label{yx_d_r_bis}
\end{equation}
This decomposition can be done for any type of generic piecewise nonlinear function $y(x)$ with discontinuities.

An algorithm for computing the describing function $F(X)$ of a piecewise nonlinear function $y(x)$ with discontinuities is reported in Algorithm~\ref{Algorithm_optimal_eff}. Let $p_i\!=\!(x_i,y_i)$, for $i\in\{1,\,2,\,\ldots,\,r\}$, denote the points where function $y(x)$ changes its slope or has a discontinuity. The algorithm takes as input vectors $\x=[x_1,\,x_2,\,\ldots,\,x_r]$ and $\y=[y_1,\,y_2,\,\ldots,\,y_r]$, defining the position of points $p_i$, and vector $\X=(0:\delta_{ X}:X_m)$, containing the desired values of amplitude $X$ of the input sinusoidal signal. The algorithm produces as output a vector $\F(\X)$ containing the samples of the resulting describing function $F(X)$ in correspondence of the points in the input vector $\X$. The Matlab code of Algorithm~\ref{Algorithm_optimal_eff} is freely available in the repository~\cite{TebaldiZanasicode2025}.

The points $p_i$ corresponding to vectors  $\x=[2,5,5,9,9,13,19]$ and $\y=[0,4,2,4,6,6,8]$ used for the example of Fig.~\ref{fig:piecewise_nonlinear} are highlighted in red in Fig.~\ref{fig:piecewise_nonlinear}(a), while the describing function $F(X)$ given by Algorithm~\ref{Algorithm_optimal_eff} is shown in Fig.~\ref{fig:piecewise_nonlinear}(c).

\begin{algorithm}[t]
\small
 \caption{Describing Function $\F(\X)$}
 \begin{algorithmic}[1]
{\color{mio_red}
 \renewcommand{\algorithmicrequire}{\textbf{Input: $\x,\y,\X$}}
 \renewcommand{\algorithmicensure}{\textbf{Output: $\F(\X)$} $\hspace{5cm}$ }
 \REQUIRE
 \ENSURE
 \STATE Initialize $\F(\X)=(0,0,\ldots,0) \in \mathbb{R}
^n$, where $n=\mbox{dim}(\X)$;
 \STATE Initialize $i=2$;
\WHILE{$i<\mbox{dim}(\x)$}
 \STATE Find set $\cS=(k,k+1,\ldots,n)$ s.t. $X_k>\x_i$;
 \IF{$x_i=x_{i+1}$}
 \STATE Compute discontinuity $Y_k=y_{i+1}-y_{i}$ as in Fig.~\ref{fig:y2xr}(a);
 \STATE Compute $\Psi(x_i,\X(\cS))$ as in \eqref{Psi_function};
 \STATE Compute $\F(\X)=\F(\X)+Y_k \Psi(x_i,\X(\cS))$ as in \eqref{FX_of_rele_with_dead_zone};
 \STATE Remove $x_i$ from $\x$; Remove $y_i$ from $\y$;
 \ENDIF
 \STATE Compute the two slopes $m_i$ and $m_{i+1}$ of the two segments $[(x_{i-1},y_{i-1}),(x_{i},y_{i})]$ and $[(x_{i},y_{i}),(x_{i+1},y_{i+1})]$;
\STATE Compute  $\Phi\!\left(\!\frac{\X(\cS)}{x_i}\!\right)$ as in \eqref{Phi_function};
  \STATE Compute $\F(\X)=\F(\X)+(m_{i+1}-m_{i}) \Phi\!\left(\!\frac{\X(\cS)}{x_i}\!\right)$ as in \eqref{FX_of_dead_zone};
\STATE Increment index $i=i+1$;
\ENDWHILE
 \RETURN $\F(\X)$
} \end{algorithmic}
\label{Algorithm_optimal_eff}
\end{algorithm}
\vspace{-2mm}

\section{Describing function $F(X)$: qualitative graphical representation}\label{Section_one_sect3}

Despite the describing function method being non-rigorous, the exact calculation of the describing function $F(X)$ usually requires complex analytical calculations. This numerical complexity greatly limits the use of this method, especially from the point of view of control education, as the compromise between complexity and precision of the results is not advantageous. On the contrary, if an approximated drawing of the describing function plot could be easily done by hand, it would enhance the interest in this method for an initial fast qualitative analysis of limit cycles.

In this section, an intuitive and fast approach for the qualitative hand-drawn graphical representation of  function $F(X)$ is therefore proposed.

\begin{Prop} \label{Prop_2}
{\bf Qualitative graphical behavior $\tilde{F}(X)$ of function $F(X)$.}
Let $y(x)$ be a
nonlinear function with discontinuities; let $X_j$, for $j\in\{1,2,\ldots,r\}$, be the values of $x$ where function $y(x)$ has a discontinuity or changes its slope from $m_{j-1}$ to $m_j$; let vectors $\overline{\X}$ and $\m$ be defined as follows $\overline{\X}=[0,X_1,\ldots,X_r]$ and $\m=[m_0,m_1,\ldots,m_r]$; let $\overline{\X}_d\in\overline{\X}$ be a subset of vector $\overline{\X}$ containing all the values $X_j$ where function $y(x)$ has a discontinuity; and let $\overline{\Y}_d$ be the vector of all the discontinuity amplitudes $Y_j$ that occur at $X=X_j\in \overline{\X}_d$. The qualitative hand-drawn graphical behavior  $\tilde{F}(X)$ of function $F(X)$ can be determined using the following rules:

1) Function $\tilde{F}(X)$ is always continuous  for $X>0$.

2)
If function $y(x)$ is discontinuous for $x=0$, then the initial value of function $\tilde{F}(X)$ is infinite:  $\tilde{F}(X)|_{X=0}=\infty$.

3)
For $X\le X_1$, the value of function $\tilde{F}(X)$  is equal to the slope $m_0$ of the first segment of function $y(x)$.

4)
 For $X\rightarrow\infty$,
the value of function $\tilde{F}(X)$  is equal to the slope $m_r$ of the last linear segment of function $y(x)$.

5) Let $\tilde{F}_j(X)$ denote the qualitative graphical behavior of function $F(X)$ within the range $X\!\in\![X_j,\,X_{j+1}]$ where function $y(x)$ has slope $m_j$. The sequence of qualitative functions  $\tilde{F}_j(X)$, for $j\in\{0,1,\ldots,r\}$,  can be plotted using Algorithm~\ref{Algorithm_sequence_Fx}. For the hand-drawing approach, note that the following rules apply:
a) $\tilde\Phi\!\left(\!\frac{X}{X_j}\!\right)$ is an exponential-like function, having a shape similar to that of function $F_d(X)$ shown in Fig.~\ref{fig:y2xd}(b): it is equal to zero until $X=X_j$, where it has null slope, and then tends to one when $X\rightarrow\infty$;
b) $\tilde\Psi\!\left(\!X,X_j\!\right)$ is an impulsive function,  having a shape similar to that of function $F_r(X)$ shown in Fig.~\ref{fig:y2xr}(b): it is equal to zero until $X=X_j$, where it has infinite slope, reaches its maximum $\frac{2}{\pi X_j}$ for $X=\sqrt{2}X_j$, and then tends to zero when $X\rightarrow\infty$.
The value of the first function  $\tilde{F}_0(X)$, for $X\!\in\![0,\,X_{1}]$, is equal to the first slope $m_0$ of function $y(x)$. The qualitative
function  $\tilde{F}_j(X)$, for $j\in\{1,2,\ldots,r\}$, starts from the value $\tilde{F}_{j-1}(X_j)$ and then tends to
the value $m_j$ for $X\rightarrow\infty$.
 If a discontinuity occurs at $X\!=\!X_j$, then
an additional term $Y_j\tilde\Psi\!\left(\!X,X_j\!\right)$ must be added to function  $\tilde{F}_j(X)$. The obtained  function  $\tilde{F}_j(X)$
 must be considered only within the range  $X\!\in\![X_j,\;X_{j+1}]$.  The value $\tilde{F}_j(X_{j+1})$  is then used
as starting point for plotting the qualitative function $\tilde{F}_{j+1}(X)$
in the next range $X\!\in\![X_{j+1},X_{j+2}]$.
$\hfill \Box$

\end{Prop}
\vspace{1mm}

 \begin{figure}[tbp]
 \centering
\setlength{\unitlength}{3.2mm}
\psset{yunit=0.9\unitlength}
\psset{xunit=0.9\unitlength}
\SpecialCoor
\begin{pspicture}(-2,-1)(26,8)
\footnotesize
\psline[linewidth=0.4pt]{->}(-0.3,0)(26,0)
\psline[linewidth=0.4pt]{->}(0,-1)(0,8)
\psline[linewidth=0.8pt]{-}(0,0)(2,0)
\rput[t](2,-0.25){$X_{1}$}
\rput[rb](1.8,0.13333){$m_{0}$}
\psline[linewidth=0.2pt,linestyle=dashed]{-}(2,0)(2,0)
\psline[linewidth=0.8pt]{-}(2,0)(7,4.5)
\rput[t](7,-0.25){$X_{2}$}
\rput[rb](4.5,2.3833){$m_{1}$}
\psline[linewidth=0.2pt,linestyle=dashed]{-}(7,0)(7,4.5)
\psline[linewidth=0.8pt]{-}(7,4.5)(20,7.2083)
\rput[t](20,-0.25){$X_{3}$}
\rput[rb](13.5,5.9875){$m_{2}$}
\psline[linewidth=0.2pt,linestyle=dashed]{-}(20,0)(20,7.2083)
\psline[linewidth=0.8pt]{-}(20,7.2083)(20,4.2083)
\psline[linewidth=0.2pt,linestyle=dashed]{-}(20,0)(20,4.2083)
\psline[linewidth=0.8pt]{-}(20,4.2083)(25,5.25)
\rput[t](25,-0.25){$X_{4}$}
\rput[rb](22.5,4.8625){$m_{3}$}
\psline[linewidth=0.2pt,linestyle=dashed]{-}(25,0)(25,5.25)
\rput[l](11.9,8){\scriptsize (a)}
\rput[t](26,-0.5){$x$}
\rput[r](-0.25,7){$y(x)$}
\end{pspicture}
\vspace{3mm}

\psset{xunit=0.9\unitlength}
\psset{yunit=10\unitlength}
\begin{pspicture}(-2,-0.1)(26,1.15)
\footnotesize
\psline[linewidth=0.4pt]{->}(-0.3,0)(26,0)
\psline[linewidth=0.4pt]{->}(0,0)(0,1.1)
\psline[linewidth=0.1pt,linestyle=dashed](2,0)(2,0)
\psline[linewidth=0.1pt,linestyle=dashed](0,0)(2,0)
\rput[r](-0.25,0){$F_1$}
\rput[t](2,-0.025){$X_1$}
\psline[linewidth=0.1pt,linestyle=dashed](0,0.9)(26,0.9)
\rput[r](-0.25,0.9){$m_1$}
\psline[linewidth=0.1pt,linestyle=dashed](7,0)(7,0.63875)
\psline[linewidth=0.1pt,linestyle=dashed](0,0.63875)(7,0.63875)
\rput[r](-0.25,0.63875){$F_2$}
\rput[t](7,-0.025){$X_2$}
\psline[linewidth=0.1pt,linestyle=dashed](0,0.20833)(26,0.20833)
\psline[linewidth=0.1pt,linestyle=dashed](20,0)(20,0.36151)
\psline[linewidth=0.1pt,linestyle=dashed](0,0.36151)(20,0.36151)
\rput[r](-0.25,0.36151){$F_3$}
\rput[t](20,-0.025){$X_3$}
\psline[linewidth=0.1pt,linestyle=dashed](0,0.20834)(26,0.20834)
\rput[r](-0.25,0.20834){$m_3$}
\psline[linewidth=1.2pt,linestyle=solid,linecolor=green](0,0)(0.13,0)(0.26,0)(0.39,0)(0.52,0)(0.65,0)(0.78,0)(0.91,0)(1.04,0)(1.17,0)(1.3,0)(1.43,0)(1.56,0)(1.69,0)(1.82,0)(1.95,0)(2.08,0.034615)(2.21,0.08552)(2.34,0.13077)(2.47,0.17126)(2.6,0.20769)(2.73,0.24066)(2.86,0.27063)(2.99,0.29799)(3.12,0.32308)(3.25,0.34615)(3.38,0.36746)(3.51,0.38718)(3.64,0.40549)(3.77,0.42255)(3.9,0.43846)(4.03,0.45335)(4.16,0.46731)(4.29,0.48042)(4.42,0.49276)(4.55,0.5044)(4.68,0.51538)(4.81,0.52578)(4.94,0.53563)(5.07,0.54497)(5.2,0.55385)(5.33,0.56229)(5.46,0.57033)(5.59,0.578)(5.72,0.58531)(5.85,0.59231)(5.98,0.599)(6.11,0.6054)(6.24,0.61154)(6.37,0.61743)(6.5,0.62308)(6.63,0.62851)(6.76,0.63373)(6.89,0.63875)(7.02,0.64235)(7.15,0.63446)(7.28,0.62685)(7.41,0.61951)(7.54,0.61242)(7.67,0.60557)(7.8,0.59895)(7.93,0.59254)(8.06,0.58635)(8.19,0.58035)(8.32,0.57453)(8.45,0.5689)(8.58,0.56344)(8.71,0.55814)(8.84,0.55299)(8.97,0.548)(9.1,0.54315)(9.23,0.53843)(9.36,0.53384)(9.49,0.52939)(9.62,0.52505)(9.75,0.52082)(9.88,0.51671)(10.01,0.51271)(10.14,0.50881)(10.27,0.505)(10.4,0.50129)(10.53,0.49768)(10.66,0.49415)(10.79,0.4907)(10.92,0.48734)(11.05,0.48406)(11.18,0.48085)(11.31,0.47772)(11.44,0.47466)(11.57,0.47167)(11.7,0.46874)(11.83,0.46588)(11.96,0.46308)(12.09,0.46034)(12.22,0.45766)(12.35,0.45504)(12.48,0.45247)(12.61,0.44995)(12.74,0.44748)(12.87,0.44507)(13,0.4427)(13.13,0.44038)(13.26,0.43811)(13.39,0.43587)(13.52,0.43369)(13.65,0.43154)(13.78,0.42943)(13.91,0.42737)(14.04,0.42534)(14.17,0.42335)(14.3,0.42139)(14.43,0.41948)(14.56,0.41759)(14.69,0.41574)(14.82,0.41392)(14.95,0.41213)(15.08,0.41037)(15.21,0.40865)(15.34,0.40695)(15.47,0.40528)(15.6,0.40364)(15.73,0.40203)(15.86,0.40044)(15.99,0.39888)(16.12,0.39734)(16.25,0.39583)(16.38,0.39434)(16.51,0.39287)(16.64,0.39143)(16.77,0.39001)(16.9,0.38862)(17.03,0.38724)(17.16,0.38588)(17.29,0.38455)(17.42,0.38323)(17.55,0.38194)(17.68,0.38066)(17.81,0.3794)(17.94,0.37816)(18.07,0.37694)(18.2,0.37574)(18.33,0.37455)(18.46,0.37338)(18.59,0.37223)(18.72,0.37109)(18.85,0.36997)(18.98,0.36886)(19.11,0.36777)(19.24,0.36669)(19.37,0.36563)(19.5,0.36458)(19.63,0.36354)(19.76,0.36252)(19.89,0.36151)(20.02,0.35184)(20.15,0.3363)(20.28,0.32723)(20.41,0.32014)(20.54,0.31416)(20.67,0.30892)(20.8,0.30423)(20.93,0.29996)(21.06,0.29604)(21.19,0.29242)(21.32,0.28904)(21.45,0.28587)(21.58,0.2829)(21.71,0.28009)(21.84,0.27743)(21.97,0.27492)(22.1,0.27252)(22.23,0.27024)(22.36,0.26807)(22.49,0.26599)(22.62,0.26401)(22.75,0.2621)(22.88,0.26028)(23.01,0.25853)(23.14,0.25684)(23.27,0.25523)(23.4,0.25367)(23.53,0.25217)(23.66,0.25072)(23.79,0.24933)(23.92,0.24798)(24.05,0.24669)(24.18,0.24543)(24.31,0.24422)(24.44,0.24305)(24.57,0.24191)(24.7,0.24081)(24.83,0.23975)(24.96,0.23872)(25.09,0.23772)(25.22,0.23676)(25.35,0.23582)(25.48,0.23491)(25.61,0.23402)(25.74,0.23317)(25.87,0.23234)(26,0.23153)
\psline[linewidth=0.3pt,linestyle=dashed,linecolor=blue](0,0)(0.13,0)(0.26,0)(0.39,0)(0.52,0)(0.65,0)(0.78,0)(0.91,0)(1.04,0)(1.17,0)(1.3,0)(1.43,0)(1.56,0)(1.69,0)(1.82,0)(1.95,0)(2.08,0.034615)(2.21,0.08552)(2.34,0.13077)(2.47,0.17126)(2.6,0.20769)(2.73,0.24066)(2.86,0.27063)(2.99,0.29799)(3.12,0.32308)(3.25,0.34615)(3.38,0.36746)(3.51,0.38718)(3.64,0.40549)(3.77,0.42255)(3.9,0.43846)(4.03,0.45335)(4.16,0.46731)(4.29,0.48042)(4.42,0.49276)(4.55,0.5044)(4.68,0.51538)(4.81,0.52578)(4.94,0.53563)(5.07,0.54497)(5.2,0.55385)(5.33,0.56229)(5.46,0.57033)(5.59,0.578)(5.72,0.58531)(5.85,0.59231)(5.98,0.599)(6.11,0.6054)(6.24,0.61154)(6.37,0.61743)(6.5,0.62308)(6.63,0.62851)(6.76,0.63373)(6.89,0.63875)(7.02,0.64359)(7.15,0.64825)(7.28,0.65275)(7.41,0.65709)(7.54,0.66127)(7.67,0.66532)(7.8,0.66923)(7.93,0.67301)(8.06,0.67667)(8.19,0.68022)(8.32,0.68365)(8.45,0.68698)(8.58,0.69021)(8.71,0.69334)(8.84,0.69638)(8.97,0.69933)(9.1,0.7022)(9.23,0.70498)(9.36,0.70769)(9.49,0.71033)(9.62,0.71289)(9.75,0.71538)(9.88,0.71781)(10.01,0.72018)(10.14,0.72249)(10.27,0.72473)(10.4,0.72692)(10.53,0.72906)(10.66,0.73114)(10.79,0.73318)(10.92,0.73516)(11.05,0.7371)(11.18,0.739)(11.31,0.74085)(11.44,0.74266)(11.57,0.74443)(11.7,0.74615)(11.83,0.74784)(11.96,0.7495)(12.09,0.75112)(12.22,0.7527)(12.35,0.75425)(12.48,0.75577)(12.61,0.75726)(12.74,0.75871)(12.87,0.76014)(13,0.76154)(13.13,0.76291)(13.26,0.76425)(13.39,0.76557)(13.52,0.76686)(13.65,0.76813)(13.78,0.76938)(13.91,0.7706)(14.04,0.77179)(14.17,0.77297)(14.3,0.77413)(14.43,0.77526)(14.56,0.77637)(14.69,0.77747)(14.82,0.77854)(14.95,0.7796)(15.08,0.78064)(15.21,0.78166)(15.34,0.78266)(15.47,0.78365)(15.6,0.78462)(15.73,0.78557)(15.86,0.78651)(15.99,0.78743)(16.12,0.78834)(16.25,0.78923)(16.38,0.79011)(16.51,0.79098)(16.64,0.79183)(16.77,0.79267)(16.9,0.79349)(17.03,0.7943)(17.16,0.7951)(17.29,0.79589)(17.42,0.79667)(17.55,0.79744)(17.68,0.79819)(17.81,0.79893)(17.94,0.79967)(18.07,0.80039)(18.2,0.8011)(18.33,0.8018)(18.46,0.80249)(18.59,0.80317)(18.72,0.80385)(18.85,0.80451)(18.98,0.80516)(19.11,0.80581)(19.24,0.80644)(19.37,0.80707)(19.5,0.80769)(19.63,0.8083)(19.76,0.80891)(19.89,0.8095)(20.02,0.81009)(20.15,0.81067)(20.28,0.81124)(20.41,0.81181)(20.54,0.81237)(20.67,0.81292)(20.8,0.81346)(20.93,0.814)(21.06,0.81453)(21.19,0.81505)(21.32,0.81557)(21.45,0.81608)(21.58,0.81659)(21.71,0.81709)(21.84,0.81758)(21.97,0.81807)(22.1,0.81855)(22.23,0.81903)(22.36,0.8195)(22.49,0.81996)(22.62,0.82042)(22.75,0.82088)(22.88,0.82133)(23.01,0.82177)(23.14,0.82221)(23.27,0.82265)(23.4,0.82308)(23.53,0.8235)(23.66,0.82392)(23.79,0.82434)(23.92,0.82475)(24.05,0.82516)(24.18,0.82556)(24.31,0.82596)(24.44,0.82635)(24.57,0.82674)(24.7,0.82713)(24.83,0.82751)(24.96,0.82788)(25.09,0.82826)(25.22,0.82863)(25.35,0.82899)(25.48,0.82936)(25.61,0.82971)(25.74,0.83007)(25.87,0.83042)(26,0.83077)
\psline[linewidth=0.3pt,linestyle=dashed,linecolor=blue](0,0)(0.13,0)(0.26,0)(0.39,0)(0.52,0)(0.65,0)(0.78,0)(0.91,0)(1.04,0)(1.17,0)(1.3,0)(1.43,0)(1.56,0)(1.69,0)(1.82,0)(1.95,0)(2.08,0.034615)(2.21,0.08552)(2.34,0.13077)(2.47,0.17126)(2.6,0.20769)(2.73,0.24066)(2.86,0.27063)(2.99,0.29799)(3.12,0.32308)(3.25,0.34615)(3.38,0.36746)(3.51,0.38718)(3.64,0.40549)(3.77,0.42255)(3.9,0.43846)(4.03,0.45335)(4.16,0.46731)(4.29,0.48042)(4.42,0.49276)(4.55,0.5044)(4.68,0.51538)(4.81,0.52578)(4.94,0.53563)(5.07,0.54497)(5.2,0.55385)(5.33,0.56229)(5.46,0.57033)(5.59,0.578)(5.72,0.58531)(5.85,0.59231)(5.98,0.599)(6.11,0.6054)(6.24,0.61154)(6.37,0.61743)(6.5,0.62308)(6.63,0.62851)(6.76,0.63373)(6.89,0.63875)(7.02,0.64235)(7.15,0.63446)(7.28,0.62685)(7.41,0.61951)(7.54,0.61242)(7.67,0.60557)(7.8,0.59895)(7.93,0.59254)(8.06,0.58635)(8.19,0.58035)(8.32,0.57453)(8.45,0.5689)(8.58,0.56344)(8.71,0.55814)(8.84,0.55299)(8.97,0.548)(9.1,0.54315)(9.23,0.53843)(9.36,0.53384)(9.49,0.52939)(9.62,0.52505)(9.75,0.52082)(9.88,0.51671)(10.01,0.51271)(10.14,0.50881)(10.27,0.505)(10.4,0.50129)(10.53,0.49768)(10.66,0.49415)(10.79,0.4907)(10.92,0.48734)(11.05,0.48406)(11.18,0.48085)(11.31,0.47772)(11.44,0.47466)(11.57,0.47167)(11.7,0.46874)(11.83,0.46588)(11.96,0.46308)(12.09,0.46034)(12.22,0.45766)(12.35,0.45504)(12.48,0.45247)(12.61,0.44995)(12.74,0.44748)(12.87,0.44507)(13,0.4427)(13.13,0.44038)(13.26,0.43811)(13.39,0.43587)(13.52,0.43369)(13.65,0.43154)(13.78,0.42943)(13.91,0.42737)(14.04,0.42534)(14.17,0.42335)(14.3,0.42139)(14.43,0.41948)(14.56,0.41759)(14.69,0.41574)(14.82,0.41392)(14.95,0.41213)(15.08,0.41037)(15.21,0.40865)(15.34,0.40695)(15.47,0.40528)(15.6,0.40364)(15.73,0.40203)(15.86,0.40044)(15.99,0.39888)(16.12,0.39734)(16.25,0.39583)(16.38,0.39434)(16.51,0.39287)(16.64,0.39143)(16.77,0.39001)(16.9,0.38862)(17.03,0.38724)(17.16,0.38588)(17.29,0.38455)(17.42,0.38323)(17.55,0.38194)(17.68,0.38066)(17.81,0.3794)(17.94,0.37816)(18.07,0.37694)(18.2,0.37574)(18.33,0.37455)(18.46,0.37338)(18.59,0.37223)(18.72,0.37109)(18.85,0.36997)(18.98,0.36886)(19.11,0.36777)(19.24,0.36669)(19.37,0.36563)(19.5,0.36458)(19.63,0.36354)(19.76,0.36252)(19.89,0.36151)(20.02,0.36052)(20.15,0.35954)(20.28,0.35857)(20.41,0.35761)(20.54,0.35667)(20.67,0.35573)(20.8,0.35481)(20.93,0.3539)(21.06,0.353)(21.19,0.35212)(21.32,0.35124)(21.45,0.35037)(21.58,0.34952)(21.71,0.34867)(21.84,0.34784)(21.97,0.34701)(22.1,0.3462)(22.23,0.34539)(22.36,0.34459)(22.49,0.3438)(22.62,0.34303)(22.75,0.34226)(22.88,0.3415)(23.01,0.34074)(23.14,0.34)(23.27,0.33926)(23.4,0.33854)(23.53,0.33782)(23.66,0.33711)(23.79,0.3364)(23.92,0.33571)(24.05,0.33502)(24.18,0.33434)(24.31,0.33366)(24.44,0.333)(24.57,0.33234)(24.7,0.33168)(24.83,0.33104)(24.96,0.3304)(25.09,0.32977)(25.22,0.32914)(25.35,0.32852)(25.48,0.32791)(25.61,0.3273)(25.74,0.3267)(25.87,0.3261)(26,0.32552)
\psline[linewidth=0.3pt,linestyle=dashed,linecolor=blue](0,0)(0.13,0)(0.26,0)(0.39,0)(0.52,0)(0.65,0)(0.78,0)(0.91,0)(1.04,0)(1.17,0)(1.3,0)(1.43,0)(1.56,0)(1.69,0)(1.82,0)(1.95,0)(2.08,0.034615)(2.21,0.08552)(2.34,0.13077)(2.47,0.17126)(2.6,0.20769)(2.73,0.24066)(2.86,0.27063)(2.99,0.29799)(3.12,0.32308)(3.25,0.34615)(3.38,0.36746)(3.51,0.38718)(3.64,0.40549)(3.77,0.42255)(3.9,0.43846)(4.03,0.45335)(4.16,0.46731)(4.29,0.48042)(4.42,0.49276)(4.55,0.5044)(4.68,0.51538)(4.81,0.52578)(4.94,0.53563)(5.07,0.54497)(5.2,0.55385)(5.33,0.56229)(5.46,0.57033)(5.59,0.578)(5.72,0.58531)(5.85,0.59231)(5.98,0.599)(6.11,0.6054)(6.24,0.61154)(6.37,0.61743)(6.5,0.62308)(6.63,0.62851)(6.76,0.63373)(6.89,0.63875)(7.02,0.64235)(7.15,0.63446)(7.28,0.62685)(7.41,0.61951)(7.54,0.61242)(7.67,0.60557)(7.8,0.59895)(7.93,0.59254)(8.06,0.58635)(8.19,0.58035)(8.32,0.57453)(8.45,0.5689)(8.58,0.56344)(8.71,0.55814)(8.84,0.55299)(8.97,0.548)(9.1,0.54315)(9.23,0.53843)(9.36,0.53384)(9.49,0.52939)(9.62,0.52505)(9.75,0.52082)(9.88,0.51671)(10.01,0.51271)(10.14,0.50881)(10.27,0.505)(10.4,0.50129)(10.53,0.49768)(10.66,0.49415)(10.79,0.4907)(10.92,0.48734)(11.05,0.48406)(11.18,0.48085)(11.31,0.47772)(11.44,0.47466)(11.57,0.47167)(11.7,0.46874)(11.83,0.46588)(11.96,0.46308)(12.09,0.46034)(12.22,0.45766)(12.35,0.45504)(12.48,0.45247)(12.61,0.44995)(12.74,0.44748)(12.87,0.44507)(13,0.4427)(13.13,0.44038)(13.26,0.43811)(13.39,0.43587)(13.52,0.43369)(13.65,0.43154)(13.78,0.42943)(13.91,0.42737)(14.04,0.42534)(14.17,0.42335)(14.3,0.42139)(14.43,0.41948)(14.56,0.41759)(14.69,0.41574)(14.82,0.41392)(14.95,0.41213)(15.08,0.41037)(15.21,0.40865)(15.34,0.40695)(15.47,0.40528)(15.6,0.40364)(15.73,0.40203)(15.86,0.40044)(15.99,0.39888)(16.12,0.39734)(16.25,0.39583)(16.38,0.39434)(16.51,0.39287)(16.64,0.39143)(16.77,0.39001)(16.9,0.38862)(17.03,0.38724)(17.16,0.38588)(17.29,0.38455)(17.42,0.38323)(17.55,0.38194)(17.68,0.38066)(17.81,0.3794)(17.94,0.37816)(18.07,0.37694)(18.2,0.37574)(18.33,0.37455)(18.46,0.37338)(18.59,0.37223)(18.72,0.37109)(18.85,0.36997)(18.98,0.36886)(19.11,0.36777)(19.24,0.36669)(19.37,0.36563)(19.5,0.36458)(19.63,0.36354)(19.76,0.36252)(19.89,0.36151)(20.02,0.35184)(20.15,0.3363)(20.28,0.32723)(20.41,0.32014)(20.54,0.31416)(20.67,0.30892)(20.8,0.30423)(20.93,0.29996)(21.06,0.29604)(21.19,0.29242)(21.32,0.28904)(21.45,0.28587)(21.58,0.2829)(21.71,0.28009)(21.84,0.27743)(21.97,0.27492)(22.1,0.27252)(22.23,0.27024)(22.36,0.26807)(22.49,0.26599)(22.62,0.26401)(22.75,0.2621)(22.88,0.26028)(23.01,0.25853)(23.14,0.25684)(23.27,0.25523)(23.4,0.25367)(23.53,0.25217)(23.66,0.25072)(23.79,0.24933)(23.92,0.24798)(24.05,0.24669)(24.18,0.24543)(24.31,0.24422)(24.44,0.24305)(24.57,0.24191)(24.7,0.24081)(24.83,0.23975)(24.96,0.23872)(25.09,0.23772)(25.22,0.23676)(25.35,0.23582)(25.48,0.23491)(25.61,0.23402)(25.74,0.23317)(25.87,0.23234)(26,0.23153)
\psline[linewidth=0.6pt,linestyle=solid,linecolor=red](0,0)(0.13,0)(0.26,0)(0.39,0)(0.52,0)(0.65,0)(0.78,0)(0.91,0)(1.04,0)(1.17,0)(1.3,0)(1.43,0)(1.56,0)(1.69,0)(1.82,0)(1.95,0)(2.08,0.008102)(2.21,0.031191)(2.34,0.058516)(2.47,0.087072)(2.6,0.11553)(2.73,0.14325)(2.86,0.16988)(2.99,0.19529)(3.12,0.21942)(3.25,0.24228)(3.38,0.2639)(3.51,0.28435)(3.64,0.30367)(3.77,0.32195)(3.9,0.33923)(4.03,0.3556)(4.16,0.37111)(4.29,0.38581)(4.42,0.39977)(4.55,0.41303)(4.68,0.42564)(4.81,0.43764)(4.94,0.44907)(5.07,0.45998)(5.2,0.47038)(5.33,0.48033)(5.46,0.48984)(5.59,0.49894)(5.72,0.50765)(5.85,0.51601)(5.98,0.52402)(6.11,0.53172)(6.24,0.53911)(6.37,0.54622)(6.5,0.55306)(6.63,0.55964)(6.76,0.56598)(6.89,0.5721)(7.02,0.57787)(7.15,0.58118)(7.28,0.58297)(7.41,0.58379)(7.54,0.58391)(7.67,0.58346)(7.8,0.58258)(7.93,0.58133)(8.06,0.57979)(8.19,0.578)(8.32,0.57601)(8.45,0.57385)(8.58,0.57156)(8.71,0.56914)(8.84,0.56664)(8.97,0.56405)(9.1,0.5614)(9.23,0.55871)(9.36,0.55597)(9.49,0.55321)(9.62,0.55043)(9.75,0.54763)(9.88,0.54483)(10.01,0.54202)(10.14,0.53922)(10.27,0.53643)(10.4,0.53364)(10.53,0.53087)(10.66,0.52812)(10.79,0.52539)(10.92,0.52267)(11.05,0.51998)(11.18,0.51732)(11.31,0.51468)(11.44,0.51206)(11.57,0.50947)(11.7,0.50691)(11.83,0.50438)(11.96,0.50188)(12.09,0.4994)(12.22,0.49696)(12.35,0.49454)(12.48,0.49215)(12.61,0.4898)(12.74,0.48747)(12.87,0.48517)(13,0.4829)(13.13,0.48066)(13.26,0.47845)(13.39,0.47627)(13.52,0.47412)(13.65,0.47199)(13.78,0.4699)(13.91,0.46783)(14.04,0.46579)(14.17,0.46377)(14.3,0.46178)(14.43,0.45982)(14.56,0.45789)(14.69,0.45598)(14.82,0.45409)(14.95,0.45223)(15.08,0.4504)(15.21,0.44859)(15.34,0.4468)(15.47,0.44504)(15.6,0.4433)(15.73,0.44158)(15.86,0.43988)(15.99,0.43821)(16.12,0.43656)(16.25,0.43493)(16.38,0.43332)(16.51,0.43173)(16.64,0.43017)(16.77,0.42862)(16.9,0.42709)(17.03,0.42558)(17.16,0.42409)(17.29,0.42262)(17.42,0.42117)(17.55,0.41974)(17.68,0.41832)(17.81,0.41692)(17.94,0.41554)(18.07,0.41418)(18.2,0.41283)(18.33,0.4115)(18.46,0.41018)(18.59,0.40888)(18.72,0.4076)(18.85,0.40633)(18.98,0.40508)(19.11,0.40384)(19.24,0.40262)(19.37,0.40141)(19.5,0.40021)(19.63,0.39903)(19.76,0.39787)(19.89,0.39671)(20.02,0.38705)(20.15,0.37136)(20.28,0.36214)(20.41,0.35491)(20.54,0.34878)(20.67,0.3434)(20.8,0.33856)(20.93,0.33415)(21.06,0.33009)(21.19,0.32632)(21.32,0.3228)(21.45,0.31949)(21.58,0.31637)(21.71,0.31343)(21.84,0.31063)(21.97,0.30797)(22.1,0.30544)(22.23,0.30302)(22.36,0.30071)(22.49,0.2985)(22.62,0.29637)(22.75,0.29433)(22.88,0.29237)(23.01,0.29049)(23.14,0.28867)(23.27,0.28692)(23.4,0.28523)(23.53,0.2836)(23.66,0.28202)(23.79,0.2805)(23.92,0.27903)(24.05,0.2776)(24.18,0.27622)(24.31,0.27488)(24.44,0.27358)(24.57,0.27232)(24.7,0.2711)(24.83,0.26991)(24.96,0.26876)(25.09,0.26763)(25.22,0.26655)(25.35,0.26549)(25.48,0.26446)(25.61,0.26345)(25.74,0.26248)(25.87,0.26153)(26,0.2606)
\rput[l](11.9,1.05){\scriptsize (b)}
\rput[t](25,-0.02){$X$}
\rput[r](-0.25,1){$F(X)$}
\rput[b](1,0.01){$\scr\tilde{F}_0(\!X\!)$}
\rput[br](4.5,0.5){$\scr\tilde{F}_1(\!X\!)$}
\rput[tr](12,0.47){$\scr\tilde{F}_2(\!X\!)$}
\rput[tr](22.2,0.28){$\scr\tilde{F}_3(\!X\!)$}
\end{pspicture}
  \setlength{\unitlength}{5.0mm}
 \psset{unit=\unitlength}
  \rput(0.83,-6.46){
  \psline[linewidth=0.42pt,fillcolor=white,fillstyle=solid](-4.5,10.15)(-1.55,10.15)(-1.55,11.85)(-4.5,11.85)(-4.5,10.15)
  \psline[linecolor=red](-4.35,11.55)(-3.4,11.55)
  \rput(-2.43,11.55){\notsotiny $F(X)$}
  \psline[linecolor=green](-4.35,11)(-3.4,11)
  \rput(-2.43,11){\notsotiny $\tilde{F}(X)$}
  \psline[linewidth=0.42pt,linecolor=blue,linestyle=dashed](-4.35,10.45)(-3.4,10.45)
  \rput(-2.43,10.45){\notsotiny $\tilde{F}_j(X)$}
  }
 \vspace{-2mm}
 \caption{(a) Piecewise nonlinear function $y(x)$.  (b) Describing function $F(X)$  of function $y(x)$,
  corresponding qualitative graphical representation $\tilde{F}(X)$,
 and exponential-like function $\tilde{F}_j(X)$ for $j \in \{0,\,1,\,2,\,3\}$.
 }
   \vspace{-3mm}
\label{y_x_c3_12_02_2025}
\end{figure}
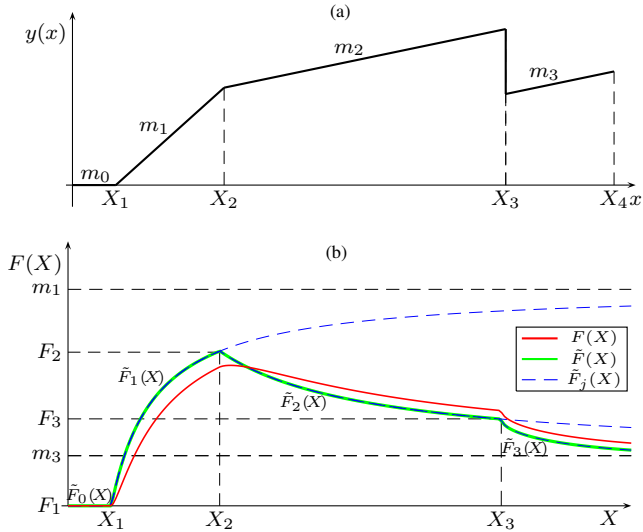

\noindent
Qualitative proofs of the aforementioned rules:

1) Function $\tilde{F}(X)$ is always continuous because the describing functions $F_d(X)$ and $F_r(X)$, composing function $F(X)$, are continuous, see Prop.~\ref{Prop_1},  Fig.~\ref{fig:y2xd} and Fig.~\ref{fig:y2xr}.

2)
 If $y(x)$ is discontinuous for $x=0$, then $\tilde{F}(X)|_{X=0}=\infty$  because, when $X\rightarrow 0$,
 the amplitude $Y_1(X)$ in \eqref{FdiX} of the first harmonic of signal $y(t)$ is finite:  $Y_1(X)=\frac{4Y_1}{\pi}$.

3)
For $X\le X_1$, function $y(x)$ is linear,
the output signal $y(t)= m_0\,x(t)$ is proportional to
$x(t)$, and  $\tilde{F}(X)=m_0$ because  the ratio $Y_1(X)/X$ in \eqref{FdiX} is equal to the slope  $m_0$.

4) When  $X\rightarrow \infty$,
 the slope $m_r$ of the last segment of function $y(x)$ becomes increasingly important for the shape of the output signal $y(t)$: when $X\rightarrow \infty$,  $y(t)\simeq m_r x(t)$ and, therefore, $\tilde{F}(X)|_{X\rightarrow\infty}=m_r$.

5) Reference is made to the nonlinear function $y(x)$ shown in  Fig.~\ref{y_x_c3_12_02_2025}(a), defined by  vectors  $\x=[2,7,20,20,25]$ and $\y=[0,    4.5,7.21,4.21,5.25]$.   The  corresponding describing function $F(X)$ (red solid line) and  qualitative graphical representation $\tilde{F}(X)$ (green solid  line) are shown  in  Fig.~\ref{y_x_c3_12_02_2025}(b).
\begin{algorithm}[t]
\small
 \caption{Qualitative Function $\tilde{\F}(\X)$}
 \begin{algorithmic}[1]
{\color{mio_red}
 \renewcommand{\algorithmicrequire}{\textbf{Input: $\X,\,\overline{\X},\,\m,\,\overline{\X}_d,\,\overline{\Y}_d$}}
 \renewcommand{\algorithmicensure}{\textbf{Output: $\tilde{\F}(X)$} $\hspace{5cm}$ }
 \REQUIRE
 \ENSURE
 \STATE Find set $\cS_0=(0,1,\ldots,n)$ s.t. $\X(\cS_0))\leq X_{1}$;
 \STATE Compute $\tilde{F}_0(\X(\cS_0))=m_0$;
 \STATE Compute $\tilde{\F}(\X(\cS_0))=\tilde{F}_0(\X(\cS_0))$;
\FOR{$j \gets 1$ to $\mbox{dim}(\overline{\X})$}
 \STATE Find set $\cS=(k,k+1,\ldots,k+n)$ s.t. $X_j< \X(\cS))\le X_{j+1}$;
 \STATE Compute $\tilde{F}_{j0}=\tilde{F}_{j-1}(X_{j})$;
 \STATE Compute $\tilde{F}_j(\X(\cS))=\tilde{F}_{j0} \!+\! (m_j\!-\!\tilde{F}_{j0})\tilde\Phi\!\left(\!\frac{\X(\cS)}{X_j}\!\right)$;
  \IF{$X_j\in \overline{\X}_d$}
 \STATE Compute $\tilde{F}_j(\X(\cS))=\tilde{F}_j(\X(\cS))+
 Y_j\tilde\Psi\!\left(\!\X(\cS),X_j\!\right)$;
 \ENDIF
 \STATE Compute $\tilde{\F}(\X(\cS))=\tilde{F}_j(\X(\cS))$;
\ENDFOR
 \RETURN $\tilde{\F}(\X)$
} \end{algorithmic}
\label{Algorithm_sequence_Fx}
\end{algorithm}
The steps for plotting the qualitative function  $\tilde{F}(X)$ are:
  \;  a)
   For $X\leq X_1$, the  qualitative function is $\tilde{F}_0(X)=0$, because the slope of the first segment of function $y(x)$ is zero.
  \; b)
  Within the range $X\in[X_1, \;X_2]$, a correct qualitative description of function $F(X)$ is described by the exponential-like function  $\tilde{F}_1(X)$ starting from point $(X_1,0)$ and reaching the value $m_1$ for $X\rightarrow\infty$, see the blue dashed line in Fig.~\ref{y_x_c3_12_02_2025}(b).
  \; c)
Within the range $X\in[X_2, \;X_3]$,  function $F(X)$ is qualitatively described by function  $\tilde{F}_2(X)$,
starting from point $(X_2,\tilde{F}_1(X_2))$ and reaching value $m_2=m_3$ for $X\!\rightarrow\!\infty$.
  \; d)
  For $X\!>\!X_3$, function $F(X)$ is qualitatively described by function  $\tilde{F}_3(X)$,
starting from point $(X_3,\tilde{F}_2(X_3))$ and reaching value $m_2\!=\!m_3$ for $X\!\rightarrow\!\infty$.

\begin{Remar}\label{remark_qualitative}

When applying the describing function method, it is an interesting and useful feature to firstly determine whether stable/unstable persistent oscillations - known as limit cycles - are present in the system in a fast way, before delving into the calculations to determine their amplitude and frequency. While the standard describing function method does not allow for such initial fast analysis, the new proposed approach for plotting the so-called \emph{qualitative describing function} does. In fact, while the qualitative function $\tilde{F}(X)$ in Fig.~\ref{y_x_c3_12_02_2025}(b) was plotted using Algorithm~\ref{Algorithm_sequence_Fx} with $\tilde\Phi\!\left(\!\frac{X}{X_j}\right)=1-\frac{X}{X_j}\!$ and $\tilde\Psi\!\left(\!X,X_j\!\right)=\Psi\!\left(\!X,X_j\!\right)$,
a qualitatively similar plot can also be effectively drawn by hand by directly applying steps from 1) to 5) in Sec. ~\ref{Section_one_sect3}. Furthermore, the qualitative describing function $\tilde{F}(X)$ captures the behavior of the actual describing function $F(X)$ with no need to perform the complex calculations in \eqref{FX_of_dead_zone}, \eqref{Psi_function} and \eqref{yx_d_r_bis}. Therefore, a simple hand-drawn sketch of the qualitative describing function $\tilde{F}(X)$ is sufficient to identify stable/unstable limit cycles in the system with sufficient detail, as discussed in the next section.
\end{Remar}

\section{Describing function: limit cycles estimation}\label{Section_one_sect4}

It is known that an estimation of the limit cycles present in the nonlinear feedback system of Fig.~\ref{nonlinear_feedback_structure} can be performed by using the following complex self-sustaining equation: $F(X)\,G(j\omega)=-1$, which equation is
is generally solved graphically by plotting both the functions $G(j\omega)$ and $-1/F(X)$ on the complex plane~\cite{desfcn2}.
The intersection points $P_i(\omega_i,X_i)$
between $G(j\omega)$ and $-1/F(X)$
provide a good estimation of the frequencies $\omega_i$ and the amplitudes $X_i$ of the limit cycles.

Due to Assumption~\ref{Assumption1}, all the nonlinear functions $y(x)$ considered in this work are characterized by a positive real describing function $F(X)$, and therefore
all the intersection points $P_i(\omega_i,X_i)$ are on the negative real semiaxis.
Furthermore, it is well known that~\cite{desfcn2}: \; 1) an intersection point $P_i(\omega_i,X_i)$
corresponds to  a stable limit cycle if, for increasing $X$,
the point $-1/F(X)$ exits from the Nyquist diagram of function  $G(j\omega)$, and to an unstable limit cycle in the opposite case;
\; 2)
the frequency and amplitude $\omega_i$ and $X_i$ of the limit cycle correspond to the frequency where function $G(j\omega_i)$ intersects the real negative semiaxis and to the value $X_i$ such that $F(X_i)=\bar{K}$, respectively, where $\bar{K}$ is the gain margin of the transfer function $G(s)$.

  \begin{figure}[tbp]
  \centering
 \setlength{\unitlength}{1.5mm}
 \psset{unit=1.0\unitlength}
 \psfrag{m2}[lb][lc][0.5]{\small$\;\;\;m_2$}
 \psfrag{m1}[lb][lb][0.5]{\small$\bar{K}_{M}$}
 \psfrag{fy}[bl][bl][0.5]{$\bar{K}^b$}
 \psfrag{fy1}[bl][bl][0.5]{$\bar{K}^c$}
 \psfrag{m0}[b][b][0.6]{\small$m_0$}
 \psfrag{mo}[l][l][0.7]{\small$m_2$}
 \psfrag{a}[tl][tl][0.5]{$(a)$}
 \psfrag{b}[tl][tl][0.5]{$(b)$}
 \psfrag{c}[tl][tl][0.5]{$(c)$}
 \psfrag{d}[t][t][0.5]{$P_2$}
 \psfrag{e}[t][t][0.4]{$P_M$}
 \psfrag{f}[tl][tl][0.5]{}
 \psfrag{g}[tl][tl][0.5]{$P_a$}
 \psfrag{1.414}[b][b][0.5]{$P_b$}
 \psfrag{h}[tl][tl][0.5]{$P_c$}
 \psfrag{1/f}[b][b][0.5]{}
 \psfrag{Diagramma di Nyquist}[b][b][0.6]{(b)}
 \psfrag{Funzione descrittiva}[b][b][0.6]{(a)}
 \psfrag{F(X)}[b][b][0.6]{ $F(X)$}
 \psfrag{X}[t][t][0.5]{$X$}
 \psfrag{X1}[lb][lb][0.5]{$X_1^b$}
 \psfrag{X2}[lb][lb][0.5]{$X_2^b$}
 \psfrag{X3}[lb][lb][0.5]{\rput(0,-2){$X_1^c$}}
 \psfrag{Real}[t][t][0.5]{Real}
 \psfrag{Imag}[b][b][0.5]{Imag}
  \includegraphics[clip,width=4.1cm]{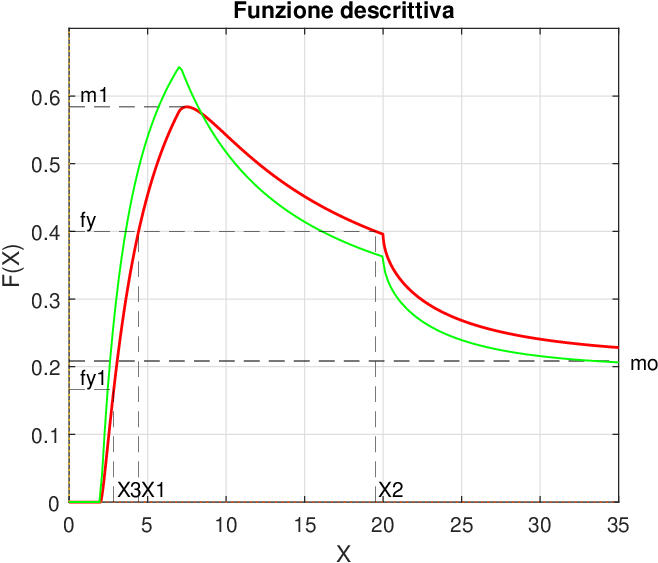}
  \hspace{2mm}
  \includegraphics[clip,width=3.8cm]{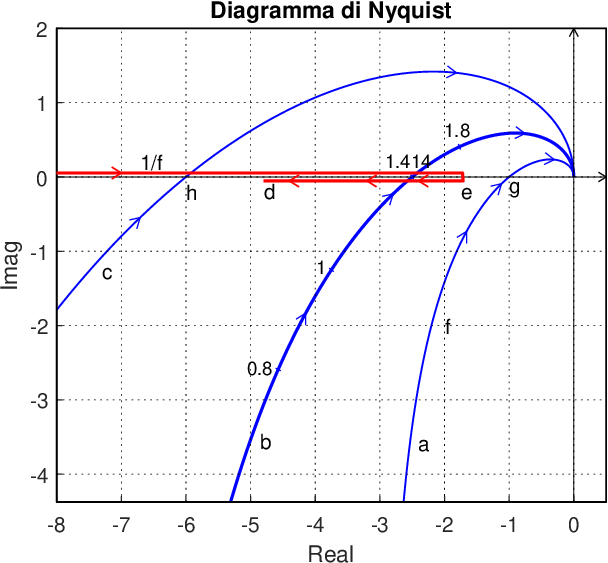}

  \setlength{\unitlength}{5.0mm}
 \psset{unit=\unitlength}
    \rput(0.8,-4.595){
\psline[linewidth=0.42pt,fillcolor=white,fillstyle=solid](-4.5,10.69)(-1.99,10.69)(-1.99,11.85)(-4.5,11.85)(-4.5,10.69)
  \psline[linecolor=red](-4.35,11.55)(-3.4,11.55)
  \rput(-2.695,11.55){\tiny $F(X)$}
  \psline[linecolor=green](-4.35,11)(-3.4,11)
  \rput(-2.695,11){\tiny $\tilde{F}(X)$}
  }
      \rput(6.02,-4.595){
\psline[linewidth=0.42pt,fillcolor=white,fillstyle=solid](-4.5,10.69)(-1.185,10.69)(-1.185,11.85)(-4.5,11.85)(-4.5,10.69)
  \psline[linecolor=red](-4.35,11)(-3.4,11)
  \rput(-2.3,11){\tiny $-1/F(X)$}
  \psline[linecolor=blue](-4.35,11.55)(-3.4,11.55)
  \rput(-2.3,11.55){\tiny $G(j\omega)$}
  }
  \vspace{-6mm}
  \caption{Case study of Sec.~\ref{first_cs_sect}. (a) Qualitative and actual describing functions $F(X)$ and $\tilde{F}(X)$ of the nonlinear function $y(x)$ in Fig.~\ref{y_x_c3_12_02_2025}(a).  (b) Graphical solution of equation $F(X)\,G(j\omega)=-1$ on the Nyquist plane. }
  \vspace{-3mm}
  \label{FX_fd_13_02_2025_C}
  \end{figure}

  The limit cycle generated by signal $y(t)$ in the  state space of system $G(s)$ can be approximately determined as follows.
Let $\cS=(\A,\,\B,\,\C,\,\D)$ be a state space realization of the
transfer function $G(s)$.
The Laplace transform $\X(s)$ of state space vector $\x(t)$ of  $\cS$ can be expressed as follows:
\begin{equation}
\X(s)=\H(s)\,Y(s),
\hspace{6mm}
\H(s)\!=\!(s\,\I -\A)\muno\B,
\label{H_di_S}
\end{equation}
where $Y(s)$ is the Laplace transform of the input $y(t)$.
 If only the first harmonic $y(t)\!=\!y_1(t)\!=\!Y_1\sin(\omega t\!+\!\varphi_1)$ is considered, the steady-state vector $\tilde{\x}(t)$ at steady-state is:
\begin{equation}
\tilde{\x}(t)\!=\!\lim_{t\rightarrow \infty} \x(t)=
\A(\omega)\sin(\omega t+\boldsymbol{\varphi}(\omega)).
\label{x_tilde}
\end{equation}
where $\A(\omega)\!=\!Y_1\left|\H(j\omega)\right|$ and $\boldsymbol{\varphi}(\omega)\!=\!\arg(\H(j\omega))+\varphi_1$.
\begin{Prop} \label{Prop_3}
 The steady-state vector $\tilde{\x}(t)$ in \eqref{x_tilde} belongs to the 2-dimensional subspace
$\cX_\omega=\mbox{span}\left(\tilde{\x}(0),\,\tilde{\x}(\frac{\pi}{2\omega})\right)$,  where $\tilde{\x}(0)=\tilde{\x}(t)|_{t=0}$ and
$\tilde{\x}(\frac{\pi}{2\omega})=\tilde{\x}(t)|_{t=\frac{\pi}{2\omega}}$.
    \end{Prop}
 \vspace{1mm}
 {\sl Proof.} From \eqref{x_tilde}, vector $\tilde{\x}(t)$ can be expressed as follows:
\[
\tilde{\x}(t)=
\sin(\omega t)
\underbrace{\A(\omega)
\sin\boldsymbol{\varphi}(\omega)}_{\tilde{\x}(0)}
+
\cos(\omega t)
\underbrace{\A(\omega)
\cos\boldsymbol{\varphi}(\omega)}_{\tilde{\x}(\frac{\pi}{2\omega})}
\]
\vspace{-1mm}
that is, for $t>0$, vector $\tilde{\x}(t)$ can be expressed as a linear combination of the two vectors $\tilde{\x}(0)$ and
$\tilde{\x}(\frac{\pi}{2\omega})$. $\hfill \Box$

  \begin{figure}[tbp]
  \centering
  \psfrag{State space trajectory}[b][t][0.6]{(a)}
  \psfrag{State x(t)}[b][b][0.6]{$x_1(t)$}
  \psfrag{Output y(t)}[t][t][0.6]{$x_2(t)=-x(t)$}
  \psfrag{X1}[b][b][0.6]{$X_1^b$}
  \psfrag{X2}[b][b][0.6]{$X_2^b$}
  \psfrag{Cu}[bl][bl][0.6]{$C_U$}
  \psfrag{Cs}[l][l][0.6]{$C_S$}
  \includegraphics[clip,width=4cm]{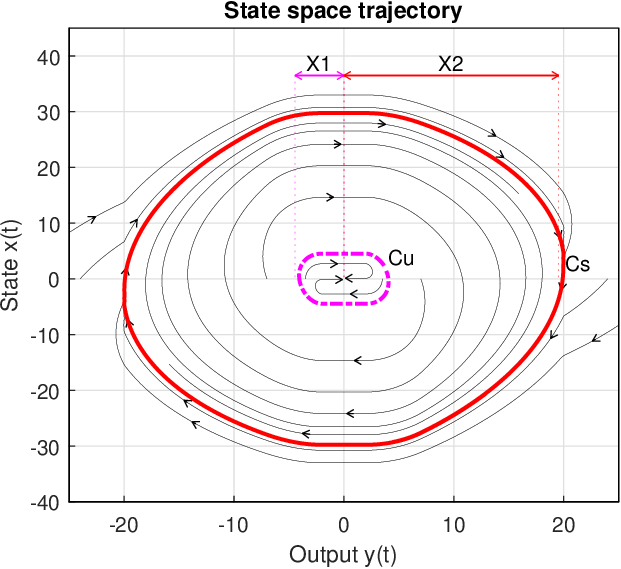}
 \hspace{2mm}
 \setlength{\unitlength}{4mm}
 \psset{unit=1.0\unitlength}
\rput(-20,0){
}
  \psfrag{X1}[b][b][0.6]{$X_1^c$}
  \psfrag{State space trajectory}[b][t][0.6]{(b)}
  \includegraphics[clip,width=4cm]{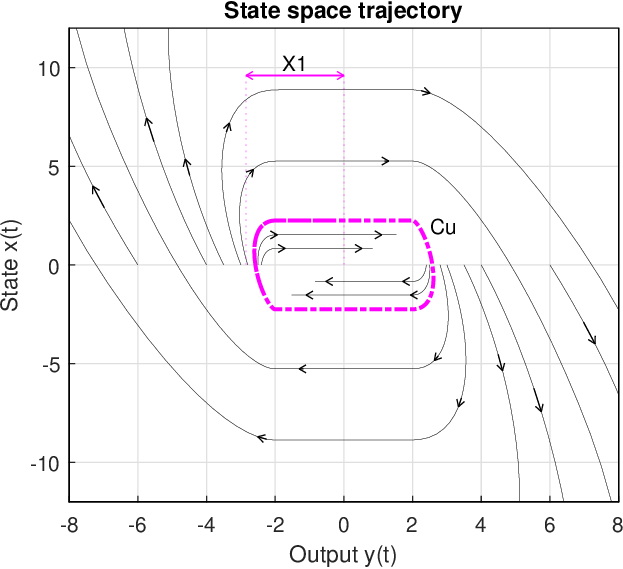}
  \vspace{-2mm}
  \caption{Case study of Sec.~\ref{first_cs_sect}. (a)
  Trajectories in case $(b)$ of Fig.~\ref{FX_fd_13_02_2025_C}(b);
  stable limit cycle  $C_S$ (red closed line) defined by point $P_{b2}=(\bar\omega,X_2^b)$, and unstable limit cycle $C_U$ (magenta dashed closed line) defined by  point $P_{b1}=(\bar\omega,X_1^b)$. (b) 
  Trajectories in case $(c)$ of Fig.~\ref{FX_fd_13_02_2025_C}(b);
  unstable limit cycle $C_U$ (magenta dashed closed line) defined by point $P_{c}=(\bar\omega,X_1^c)$.}
  \vspace{-3mm}
  \label{Ciclo_Limite_ii_5_Fig_4}
  \end{figure}
%

\subsection{First Case Study}\label{first_cs_sect}
Reference is made to the feedback system of Fig.~\ref{nonlinear_feedback_structure}, where
$G(s)=\frac{k\,(2-s)}{s(s+1)}$
 and $y(x)$ is the nonlinear function defined in Fig.~\ref{y_x_c3_12_02_2025}(a).
The Nyquist  diagrams of function $G(s)$ for $k=1$,  $k=2.5$ and  $k=6$ are shown in Fig.~\ref{FX_fd_13_02_2025_C}(b), see
labels $(a)$,  $(b)$ and  $(c)$, respectively. The intersection points of function $G(s)$ with the negative real semiaxis  are denoted by $\ts P_a\!=\!-\frac{1}{\bar{K}^a}$,  $P_b\!=\!-\frac{1}{\bar{K}^b}$ and  $P_c\!=\!-\frac{1}{\bar{K}^c}$, and are characterized by the same frequency $\bar\omega=\sqrt{2}$
and by the gain margins $\bar{K}^a=1$,  $\bar{K}^b=0.4$ and  $\bar{K}^c=0.166$.
The actual and qualitative describing functions $F(X)$ and $\tilde{F}(X)$
are shown in Fig.~\ref{FX_fd_13_02_2025_C}(a). The points $P_2$ and $P_M$ in Fig.~\ref{FX_fd_13_02_2025_C}(b) are defined as follows:
$\ts P_2\!=\!-\frac{1}{m_2}$ and $\ts P_M\!=\!-\frac{1}{\bar{K}_M}$.
The points $P_a$, $P_c$ and $P_c$ correspond to the following  three working conditions:
\; a)
$\bar{K}^a>\bar{K}_M$: function $-\frac{1}{F(X)}$ is outside the Nyquist diagram of function $G(j\omega)$, therefore the origin of the feedback system is globally asymptotically stable.
\; b)
$\bar{K}_M\!>\!\bar{K}^b\!>\!m_2$:  function $-\frac{1}{F(X)}$ intersects $G(j\omega)$ in two points: $P_{b1}\!=\!(\bar\omega,X_1^b)$ and $P_{b2}\!=\!(\bar\omega,X_2^b)$. The state space  trajectories in the vicinity of the origin are shown in Fig.~\ref{Ciclo_Limite_ii_5_Fig_4}(a), where $x(t)=-x_2(t)$ is the output of system $G(s)$.
The simulated amplitudes $X_1^b$ and $X_2^b$ of Fig.~\ref{Ciclo_Limite_ii_5_Fig_4}(a) are quite similar to
the amplitudes $X_1^b$ and $X_2^b$ in the analysis of  Fig.~\ref{FX_fd_13_02_2025_C}(a) when $F(X)=\bar{K}^b$.
 \; c)  $\bar{K}^c<m_2$:  function $-\frac{1}{F(X)}$ intersects $G(j\omega)$ in point $P_{c}=(\bar\omega,X_1^c)$, corresponding to the unstable limit cycle $C_U$ of Fig.~\ref{Ciclo_Limite_ii_5_Fig_4}(b): depending on the initial condition $(y_0,x_0)$, the system trajectories either tend to the origin or to infinity.

Note that the actual and qualitative describing functions $F(X)$ and $\tilde{F}(X)$ in Fig.~\ref{FX_fd_13_02_2025_C}(a) are qualitatively similar, thus the same conclusions as the aforementioned analysis in terms of existence of stable/unstable limit cycles in the feedback system can be achieved using the proposed qualitative describing $\tilde{F}(X)$, see Remark~\ref{remark_qualitative}. Furthermore, a very similar hand-drawn sketch of the qualitative describing function $\tilde{F}(X)$ in Fig.~\ref{FX_fd_13_02_2025_C}(a), useful for control education, can be achieved using the proposed steps from 1) to 5) in Sec. ~\ref{Section_one_sect3}.


\subsection{Second Case Study}\label{second_cs_sect}

\begin{figure}[tbp] 
    \centering
\setlength{\unitlength}{1.8mm}
\psset{unit=\unitlength}
\psset{yunit=1.4\unitlength}
\SpecialCoor
\begin{pspicture}(-1.5,-1)(20,14)
\tiny
\rput[c](10,12){(a)}
\rput[t](20,-0.5){ $x$}
\rput[r](-0.25,12){$y(x)$}
\psline[linewidth=0.4pt]{->}(-1,0)(20,0)
\psline[linewidth=0.4pt]{->}(0,-1)(0,13)
\psline[linewidth=0.8pt]{-}(0,0)(3,3)
\rput[t](3,-0.25){$3$}
\psline[linewidth=0.2pt,linestyle=dashed]{-}(3,0)(3,3)
\psline[linewidth=0.2pt,linestyle=dashed]{-}(0,3)(3,3)
\psline[linewidth=0.8pt]{-}(3,3)(6,3)
\rput[t](6,-0.25){$6$}
\psline[linewidth=0.2pt,linestyle=dashed]{-}(6,0)(6,3)
\rput[r](-0.25,3){$3$}
\psline[linewidth=0.2pt,linestyle=dashed]{-}(0,3)(6,3)
\psline[linewidth=0.8pt]{-}(6,3)(10,10)
\rput[t](10,-0.25){$10$}
\psline[linewidth=0.2pt,linestyle=dashed]{-}(10,0)(10,10)
\rput[r](-0.25,10){$10$}
\psline[linewidth=0.2pt,linestyle=dashed]{-}(0,10)(10,10)
\psline[linewidth=0.8pt]{-}(10,10)(19,10)
\rput[t](19,-0.25){$19$}
\psline[linewidth=0.2pt,linestyle=dashed]{-}(19,0)(19,10)
\rput[r](-0.25,10){$10$}
\psline[linewidth=0.2pt,linestyle=dashed]{-}(0,10)(19,10)
\end{pspicture}
\hspace{2mm}
  \psfrag{State space trajectory}[b][t][0.5]{(b)}
  \psfrag{State x(t)}[b][b][0.6]{$x_1(t)$}
  \psfrag{Output y(t)}[t][t][0.5]{$x_2(t)=-x(t)$}
  \psfrag{x1(t)}[b][b][0.5]{$x_1(t)$}
  \psfrag{x2(t)}[b][b][0.5]{$x_2(t)$}
  \psfrag{Cu}[bl][bl][0.6]{$C_U$}
  \psfrag{Cs}[tr][tr][0.6]{$C_S$}
  \includegraphics[clip,width=4.2cm]{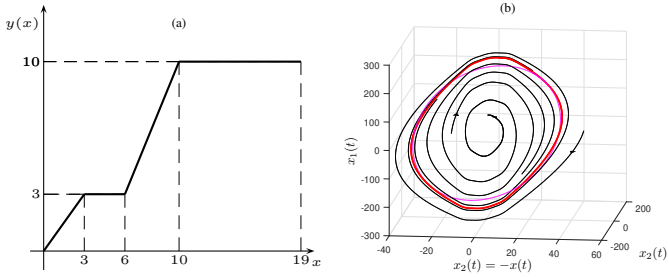}    
  \vspace{-1.68mm}
  \caption{Case study of Sec.~\ref{second_cs_sect}.  (a) Nonlinear function $y(x)$. b) State space trajectories in case $(c)$ of Fig.~\ref{FX_fd_17_02_2025_C}(b).
  }
  \label{FX_fd_18_02_2025_C}
  \vspace{-0.88mm}
  \end{figure}

\begin{figure}[tbp] \centering
  \includegraphics[clip,width=\columnwidth]{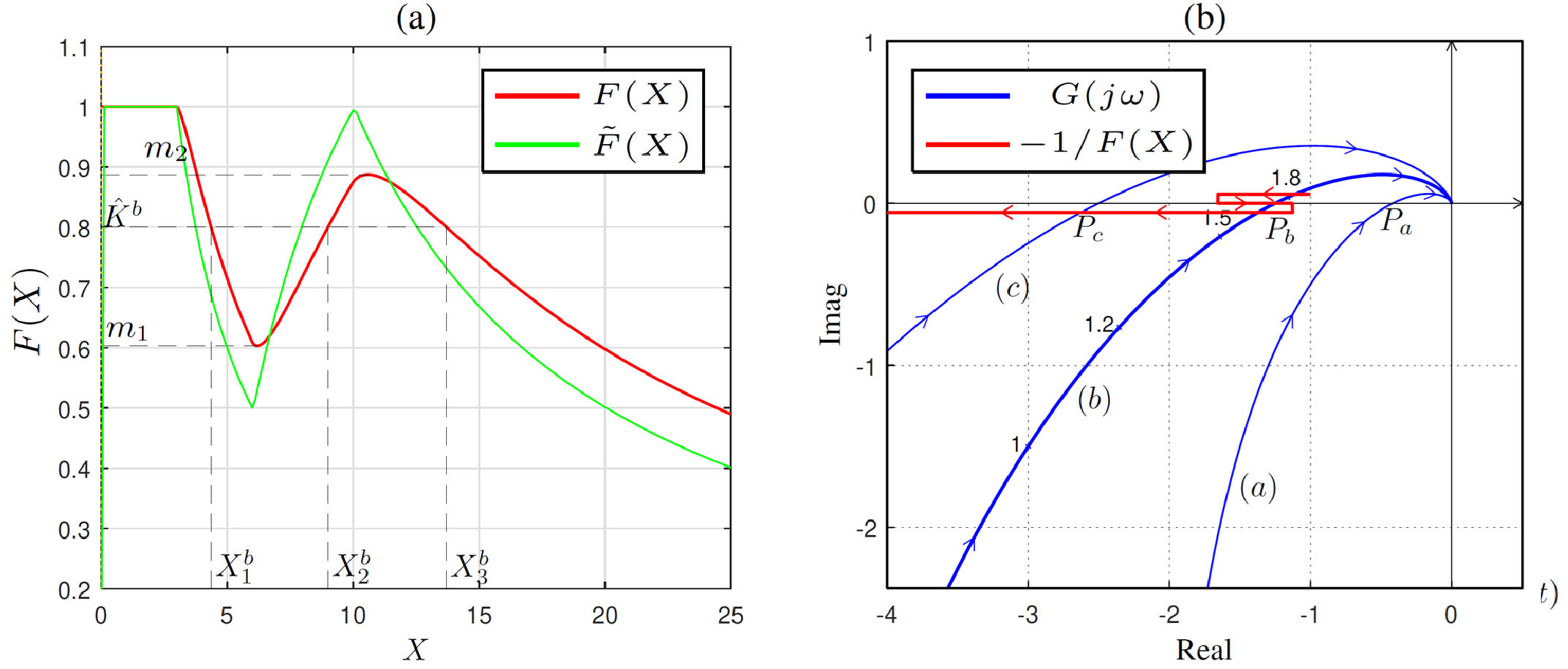}
  \vspace{-6.8mm}
  \caption{Case study of Sec.~\ref{second_cs_sect}. (a) Qualitative and actual describing functions $F(X)$ and $\tilde{F}(X)$ of the nonlinear function $y(x)$ in Fig.~\ref{FX_fd_18_02_2025_C}(a).  (b) Graphical solution of equation $F(X)\,G(j\omega)=-1$ on the Nyquist plane.
  }
  \vspace{-2.8mm}
  \label{FX_fd_17_02_2025_C}
  \end{figure}

\begin{figure}[tp]
    \centering
  \psfrag{State space trajectory}[b][t][0.5]{}
  \psfrag{State x(t)}[b][b][0.8]{$x_1(t)$}
  \psfrag{Output y(t)}[t][t][0.8]{$x_2(t)=-x(t)$}
  \psfrag{x1(t)}[b][b][0.8]{$x_1(t)$}
  \psfrag{x2(t)}[b][b][0.8]{$x_2(t)$}
  \psfrag{Cu}[bl][bl][0.8]{$C_U$}
  \psfrag{Cs}[tr][tr][0.8]{$C_S$}
  \includegraphics[clip,width=6.28cm]{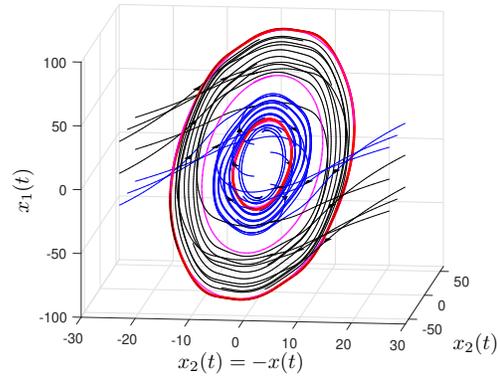}
      \vspace{-2.2mm}
  \caption{Case study of Sec.~\ref{second_cs_sect}:
  trajectories in case $(b)$ of Fig.~\ref{FX_fd_17_02_2025_C}(b).
  }
  \vspace{-4.2mm}
  \label{FX_fd_18_02_2025_C_3D}
  \end{figure}

Let us consider the feedback system in Fig.~\ref{nonlinear_feedback_structure} where
$G(s)=\frac{k}{s(s+1)(s+3)}$
 and $y(x)$ is the nonlinear function defined in Fig.~\ref{FX_fd_18_02_2025_C}(a).
The Nyquist  diagrams of function $G(s)$ for $k=5$,  $k=15$ and  $k=30$ are shown in Fig.~\ref{FX_fd_17_02_2025_C}(b), see
labels $(a)$,  $(b)$ and  $(c)$, respectively. The intersection points $\ts P_a=-\frac{1}{\bar{K}^a}$,
$ P_b=-\frac{1}{\bar{K}^b} $ and
$ P_c=-\frac{1}{\bar{K}^c}$ in Fig.~\ref{FX_fd_17_02_2025_C}(b) are characterized by the same frequency $\bar\omega=\sqrt{3}$ and gain margins $\bar{K}^a=2.4$,  $\bar{K}^b=0.8$ and  $\bar{K}^c=0.4$.
The actual and qualitative describing functions $F(X)$ and $\tilde{F}(X)$
are shown in Fig.~\ref{FX_fd_17_02_2025_C}(a).
Three cases are considered:
 \,{1)}
 Point $P_a$, which does not intersect function $-1/F(X)$, corresponds to the  global asymptotic stability of the origin.
   \,{2)}
The intersection point $P_c$ corresponds to the presence of a stable limit cycle, see Fig.~\ref{FX_fd_18_02_2025_C}(b):
 all the trajectories in the 3D space (black lines) tend to the stable limit cycle (red line), and the magenta line denotes the limit cycle estimation  $\tilde{\x}(t)\in\cX_\omega$,  defined in  Prop.~\ref{Prop_3}.
  \,{3)}
 Point $P_b$, which intersects function $-1/F(X)$ three times, corresponds to the presence of three limit cycles, see Fig.~\ref{FX_fd_18_02_2025_C_3D}: the two solid red lines denote the two stable limit cycles. The black and blue state space trajectories tend to the outer and inner stable limit cycles, respectively.  The three magenta lines in Fig.~\ref{FX_fd_18_02_2025_C_3D} denote the three limit cycle estimations  $\tilde{\x}_i(t)\in\cX_\omega$ described  in  Prop.~\ref{Prop_3}.

Even in this case, by looking at the actual and qualitative describing functions $F(X)$ and $\tilde{F}(X)$ in Fig.~\ref{FX_fd_17_02_2025_C}(a), it can be concluded that Remark~\ref{remark_qualitative} hods true, as the same conclusion reported at the end of Sec.~\ref{first_cs_sect} also apply here.

\section{Conclusions}\label{conclusions_sect}
This paper has introduced a new method for the qualitative plotting the describing function of piecewise nonlinearities involving discontinuities which, unlike the standard approach,
effectively leads to a fast hand-drawing of the qualitative describing function by following the proposed rules, starting from the characteristics of the considered nonlinearity. The analysis of limit cycles in two numerical case studies has then been performed, showing that the same qualitative results can be obtained using the standard exact plotting and the proposed qualitative hand-drawn plotting of the describing function.

\bibliographystyle{IEEEtran}

\bibliography{references}

\end{document}

%% file: mathbase.tex
\newcommand{\muno}{^{\mbox{\tiny -1}}}

\newcommand{\ts}{\textstyle}
\newcommand{\scr}{\scriptstyle}

\def\m{{\bf m}}   
   
   \def\x{{\bf x}}
\def\y{{\bf y}} 

\def\A{{\bf A}} \def\B{{\bf B}} \def\C{{\bf C}} \def\D{{\bf D}}
 \def\F{{\bf F}}  \def\H{{\bf H}}
\def\I{{\bf I}}

   \def\X{{\bf X}}
\def\Y{{\bf Y}} 

  \def\cS{{\cal S}} 
   \def\cX{{\cal X}}

